\documentclass{amsart} %{amsart}%{article}%{elsart}
\usepackage{graphicx}
\usepackage{amssymb, amsmath, sidecap}
\usepackage{amsfonts}
\usepackage{amssymb}%,showkeys}
\usepackage{float}

\newcommand{\R}{\mathbb R}

\def\la{\label}

\def\bt{\begin{thm}}
\def\et{\end{thm}}
\def\bl{\begin{lem}}
\def\el{\end{lem}}
\def\bd{\begin{defi}}
\def\ed{\end{defi}}
\def\bc{\begin{cor}}
\def\ec{\end{cor}}
\def\bp{\begin{proof}}
\def\ep{\end{proof}}
\def\br{\begin{rem}}
\def\er{\end{rem}}

\newtheorem{thm}{Theorem}[section]
\newtheorem{lem}{Lemma}[section]
\newtheorem{defi}{Definition}[section]

\newtheorem{rem}{Remark}[section]
\newtheorem{cor}{Corollary}[section]

\numberwithin{equation}{section}
\numberwithin{figure}{section}

\begin{document}
\title[Theory of Dark matter and Dark Energy]{Gravitational Field Equations  and \\ Theory of Dark Matter and Dark Energy}
\author[Ma]{Tian Ma}
\address[TM]{Department of Mathematics, Sichuan University,
Chengdu, P. R. China}

\author[Wang]{Shouhong Wang}
\address[SW]{Department of Mathematics,
Indiana University, Bloomington, IN 47405}
\email{showang@indiana.edu, http://www.indiana.edu/~fluid}

\thanks{The work was supported in part by the
Office of Naval Research and by the National Science Foundation.}


\begin{abstract} The main objective of this article is to derive a new set of gravitational field equations and to establish a new unified theory for dark energy and dark matter. The new gravitational field equations with scalar potential $\varphi$ are derived  using the Einstein-Hilbert functional, and the scalar potential $\varphi$ is a natural outcome of the divergence-free constraint of the variational elements. 
Gravitation is now described by the Riemannian metric
$g_{ij}$,  the scalar potential $\varphi$  and their interactions, unified by the new gravitational field equations.
Associated with  the scalar potential $\varphi$ is the scalar potential energy density $\frac{c^4}{8\pi G} \Phi=\frac{c^4}{8\pi G} g^{ij}D_iD_j \varphi$, which  represents a new type of  energy  caused by the non-uniform distribution of matter in the universe. 
The negative part of this potential energy density  produces attraction, and the positive part produces repelling force.   This potential energy density  is conserved with mean zero: $\int_M \Phi dM=0$.
The sum of this new potential energy density 
$\frac{c^4}{8\pi G} \Phi$ and the coupling energy 
 between the energy-momentum tensor $T_{ij}$ and the scalar potential field $\varphi$ gives rise to a new unified theory for dark matter and dark energy: 
The negative part of 
this sum represents the dark matter, which produces attraction, 
and the positive part represents the dark energy, which drives the acceleration of expanding galaxies.
In addition, the scalar curvature of space-time obeys $R=\frac{8\pi G}{c^4} T + \Phi$.
Furthermore, the new field equations resolve a few difficulties encountered by the classical Einstein field equations.
\end{abstract}
\maketitle
\tableofcontents

\section{Introduction  and Summary}
The main aim of this article is an attempt to derive a new theory for dark matter and dark energy, and to derive a new set of gravitational field equations. 

The primary motivation of this study is the great mystery of the dark matter and dark energy. The natural  starting point for this study is to fundamentally examine the Einstein field equations, given as follows:
\begin{equation}\la{1.1}
R_{ij}-\frac{1}{2}g_{ij}R =-\frac{8\pi
G}{c^4}T_{ij}, 
\end{equation}
where $R_{ij}$ is the Ricci curvature tensor, $R$  is the scalar curvature, $g_{ij}$ is the Riemannian metric of the space-time, and $T_{ij}$  is the energy-momentum tensor of matter; see among many others \cite{atwater}.
The Einstein equations can also be derived using the Principle of Lagrangian Dynamics to the  Einstein-Hilbert functional:
\begin{equation}
F(g_{ij})=\int_M R \sqrt{-g}dx, \la{1.2}
\end{equation}
whose Euler-Lagrangian is exactly $R_{ij}-\frac12 g_{ij}R$, which is the left hand side of the Einstein field equations (\ref{1.1}). 
It is postulated that this Euler-Lagrangian is balanced by the symmetric  energy-momentum tensor of matter, $T_{ij}$, leading to the  Einstein field equations (\ref{1.1}). The Bianchi identity implies that the left hand side of the Einstein equations is divergence-free, and it is then postulated and widely accepted that the energy-momentum tensor of matter $T_{ij}$  is divergence-free as well. 

However, there are a number of difficulties for the Einstein field equations:

First, the Einstein field equations failed to explain the dark matter and dark energy, and  the equations are inconsistent with the accelerating expansion of the galaxies. In spite of many attempts  to modify the Einstein gravitational field equation to derive a consistent theory for the dark energy,  the mystery remains. 

Second, we can prove that there is no solution for the Einstein field equations for the spherically symmetric case with cosmic microwave background (CMB). One needs clearly to resolve this inconsistency caused by the  non-existence of solutions. 

Third, from the Einstein equations (\ref{1.1}), it is clear that 
\begin{equation}
\la{1.1-1} R=\frac{4\pi G}{c^4} T,
\end{equation}
where  $T=g^{ij}T_{ij}$ is the energy-momentum  density. A direct consequence of this formula is  that the discontinuities of $T$ give rise to the same discontinuities of the curvature and the discontinuities of space-time. This is certainly an inconsistency which needs to be resolved.

Fourth, it has been observed that the universe is highly non-homogeneous as indicated by e.g. the "Great Walls", filaments and voids. However, the Einstein equations do not appear to offer a good explanation of this inhomogeneity. 

These observations strongly suggest that  further fundamental level  examinations of the Einstein equations are  inevitably necessary.  It is clear that any modification of the Einstein field equations should obey three basic principles: 
\begin{itemize}
\item the principle of equivalence, 
\item the principle of general relativity, and
\item the principle of Lagrangian dynamics. 
\end{itemize}
The first two principles tell us
that the spatial and temporal world is a 4-dimensional Riemannian
manifold $(M,g_{ij})$,  where the metric $\{g_{ij}\}$ represents
gravitational potential, and the third principle determines that  the Riemannian metric 
$\{g_{ij}\}$ is an extremum point of  the Lagrangian action. 
There is no doubt that the  most natural Lagrangian in this case is  the
Einstein-Hilbert functional as explained in many classical texts of general relativity.

The key observation for our study  is a well-known fact that  the Riemannian metric $g_{ij}$  is divergence-free. This suggests two important postulates for deriving a new set of gravitational field equations:

\begin{itemize}
\item The energy-momentum tensor  $T_{ij}$ of matter need not to be divergence-free due to the presence of dark energy and dark matter; and 

\item The field equations obey the Euler-Lagrange equation of the Einstein-Hilbert functional under the natural divergence-free constraint, with divergence defined at the extremum Riemannian metric $g$:
$$D^k_g X_{kl}=0, \qquad X_{kl}=X_{lk}.$$
Here $D^i_g$ is the contra-variant derivative with respect to the extremum point $g$, and $X_{ij}$  are the variational elements. Namely, for any $X=\{X_{ij}\}$  with $D^i_gX_{ij}=0$, 
$$\lim_{\lambda \to 0}\frac{1}{\lambda}[ F(g_{ij}+ \lambda X_{ij}) -F(g_{ij})]=(\delta F(g_{ij}), X)=0. $$
\end{itemize}

For this purpose, an important part of this article is  to drive an orthogonal decomposition theorem of tensors on Riemannian manifolds, which we shall explain further in the last part of this Introduction. 

Under these two postulates, using the orthogonal decomposition theorem of tensors,  we derive the following new set of gravitational field equations  with scalar potential:
\begin{equation}
R_{ij}-\frac{1}{2}g_{ij}R=-\frac{8\pi G}{c^4}T_{ij}- D_iD_j\varphi, \la{1.3}
\end{equation} 
where the scalar function $\varphi:M\to \R$ is called
the scalar potential. 

The corresponding conservations of mass, energy and momentum are then replaced by 
\begin{equation}
{\rm div}\ (D_iD_j\varphi  + \frac{8\pi
G}{c^4}T_{ij})=0, \la{1.4}
\end{equation}
and the energy-momentum density $T=g^{ij} T_{ij}$  and the scalar potential energy density  $\frac{c^4}{8\pi G}\Phi= \frac{c^4}{8\pi G}g^{ij}D_iD_j\varphi$ satisfy
\begin{align}
& R=\frac{8\pi G}{c^4}T +  \Phi, \la{1.5}\\
& \int_M \Phi \sqrt{-g} dx =0.\la{1.6}
\end{align}

The scalar potential energy density $\frac{c^4}{8\pi G}\Phi$  has a number  of  important physical properties:

\begin{itemize}

\item[1.] Gravitation is now described by the Riemannian metric
$g_{ij}$,  the scalar potential $\varphi$  and their interactions, unified by the new gravitational field equations (\ref{1.3}).

\item[2.] This scalar potential energy density $\frac{c^4}{8\pi G}\Phi$ represents a new type of  energy/force  caused by the non-uniform distribution of matter in the universe. 
This scalar potential energy density varies as  the galaxies move and matter of the universe redistributes. Like gravity, it affects every part of the universe as a field. 

\item[3.]
This scalar potential energy density $\frac{c^4}{8\pi G}\Phi$  consists of both positive and negative energies.  
The negative part of this potential energy density  produces attraction, and the positive part produces repelling force. 
The conservation law (\ref{1.6}) amounts to saying that the 
the total  scalar potential energy density  is conserved. 

%  The negative part of this quantity $\Phi$ represents the dark matter, which produces attraction, and the positive part represents the dark energy, which drives the acceleration of expanding galaxies. 

\item[4.]  
The sum  of this new potential energy density $\frac{c^4}{8\pi G} \Phi$ and the coupling energy 
between the energy-momentum tensor $T_{ij}$ and the scalar potential field $\varphi$,  as described e.g. by the second term in the right-hand side of (\ref{1.9-1}),  gives rise to a new unified theory for dark matter and dark energy: 
The negative part of $\varepsilon$  represents the dark matter, which produces attraction, and the positive part represents the dark energy, which drives the acceleration of expanding galaxies.

\item[5.] The scalar curvature of space-time obeys (\ref{1.5}). Consequently, when there is no normal matter present (with $T=0$), the curvature $R$ of space-time is balanced by $R=\Phi$. Therefore, there is no real vacuum in the universe.

\item[6.] The universe with uniform distributed matter leads to identically zero scalar potential energy, and  is unstable. It is this instability that leads to the existence of the dark matter and dark energy, and consequently the high non-homogeneity of the universe. 
%The theory strongly suggest that as soon  as the universe form after the Big Bang, 
%this scalar potential energy field, and consequently the dark matter and dark energy,  
%are present in the entire universe, filling the entire  empty space. In a sense, 
%vacuum space is no longer in existence.
\end{itemize}

Hereafter, we further explore  a  few direct consequences of the above new gravitational field equations. 

First, the new field equations  are consistent with the spherically symmetric case with cosmic microwave background (CMB). Namely, the existence of solutions in this case can be  proved.

Second, 
%consider the case where the matter is concentrated at a single point given by a delta function, classically, the curvature would be infinity at this point and becomes zero everywhere different from the point source, according to the classical field equations. 
our new theory suggests that the curvature $R$ is always balanced by  $\Phi$ in the entire space-time by (\ref{1.5}), and the space-time is no longer flat. Namely the entire space-time is also curved and is filled with dark energy and dark matter. In particular, the discontinuities of $R$ induced by the discontinuities of the energy-momentum density $T$,  dictated by the Einstein field equations, are no longer present thanks to the balance of  $ \Phi$.

Third, this scalar potential energy density should be viewed  as  the main cause for the non-homogeneous distribution of the matter/galaxies in the universe, as the dark matter (negative scalar potential energy) attracts and dark energy (positive scalar potential energy) repels different galaxies; see (\ref{1.9-1}) below.

Fourth, to further explain the dark matter and dark energy phenomena, we consider a  central matter field with  total mass $M$ and radius $r_0$ and spherical symmetry. 
With spherical coordinates,   the corresponding Riemannian metric must be of the following form: 
\begin{equation}
ds^2=-e^{u} c^2dt^2+e^v dr^2+r^2(d\theta^2+\sin^2\theta
d\varphi^2), 
\end{equation}
where $u=u(r)$ and $v=v(r)$  are functions of the radial distance. 
With  the new field equations, the  force exerted on an object with mass $m$ is given by 
\begin{equation}
F={mMG} \left[  
-\frac{1}{r^2} - \frac1{\delta} \left(2+ \frac{\delta}{r} \right) \varphi' + \frac{Rr}{\delta}\right],  \qquad R =\Phi \qquad \text{ for }r > r_0.\la{1.9-1}
\end{equation}
where $\delta=2GM/c^2$, $R$  is the scalar curvature, and $\varphi$ is the scalar potential.
The first term is the classical Newton gravitation, the second term is the coupling interaction between matter and the scalar potential $\varphi$, and the third term is the interaction generated by the scalar potential energy density $\frac{c^4}{8\pi G}\Phi$ as indicated in (\ref{1.5}) ($R=\Phi$ for $r>r_0$).
In this formula, the negative and positive values of each term represent respectively the attracting and repelling forces. 
It is then  clear that the combined effect of the second and third terms in the above formula represent the dark matter, dark energy  and their interactions with normal matter. 

Also, importantly, this formula is  a direct representation of the Einstein's equivalence principle. Namely,  the curvature of space-time induces interaction forces between matter.

In addition, one can derive a more detailed version  of the above formula:
\begin{equation}
F={mMG} \left[  
-\frac{1}{r^2} +  \left(2+ \frac{\delta}{r} \right) \varepsilon r^2 + \frac{Rr}{\delta} + 
\frac{1}{\delta} \left(2+ \frac{\delta}{r} \right) r^2 \int r^{-2} R dr
\right],\la{f0}
\end{equation}
where $\varepsilon >0$.  
The conservation law (\ref{1.6}) of $\Phi$ suggests that $R$ behaviors as $r^{-2}$  for  $r$ sufficiently large. Consequently the second term in the right hand side of (\ref{f0}) must dominate and be positive, indicating the existence of dark energy.

In fact, the above formula can be further simplified to derive the following approximate formula:
\begin{align}
& F=mMG\left[-\frac{1}{r^2} -\frac{k_0}{r} + k_1 r  \right], \la{1.9}\\
& k_0=4  \times 10^{-18} km^{-1}, \qquad k_1=10^{-57} km^{-3}.\la{1.10}
\end{align}
Again, in (\ref{1.9}), the first term represents the Newton gravitation, the attracting second term stands for dark matter and the repelling third term is the dark energy.

\medskip

The mathematical part of this article is devoted to a rigorous derivation of the new gravitational field equations.

First, as mentioned earlier, the field equations  obey the Euler-Lagrange equation of the Einstein-Hilbert functional under the natural divergence-free constraint, with divergence defined at the extremum Riemannian metric $g$:
\begin{equation}
(\delta F(g_{ij}), X)=0 \qquad \forall \ X=\{X_{ij}\} \text{ with } D^i_gX_{ij}=0, X_{ij}=X_{ji}. \la{el}
\end{equation}
 As the variational elements $X$  are divergence-free, (\ref{el}) does not imply $\delta F(g_{ij})=0$, which is the classical Einstein equations. In fact, (\ref{el}) amounts to saying that $\delta F(g_{ij})$  is orthogonal to all divergence-free tensor fields $X$. 
 
Hence we need to decompose general tensor fields on Riemannian manifolds into divergence-free and gradient parts.  For this purpose, an orthogonal decomposition theorem is derived in Theorem~\ref{t3.1}. 
In  particular, given an $(r, s)$ tensor field $u \in L^2(T^r_sM)$, we have 
\begin{equation}
\la{rsd}
u=\nabla \psi + v, \qquad \text{div } v=0, \qquad \psi \in H^1(T^r_{s-1}).
\end{equation}
The gradient part is acting on an  $(r, s-1)$ tensor field $\psi$.

Second, restricting to a $(0, 2)$ symmetric tensor field $u$, the gradient part in the above decomposition is 
given by 
$$\nabla \psi \qquad \text{ with } \psi =  \{\psi_i\} \in H^1(T^0_1M).$$
Then using symmetry, we show in Theorem~\ref{t3.4} that this $(0,1)$ tensor $\psi$  can be uniquely determined, up to addition to  constants,  by the gradient of a scalar field $\varphi$:
$$\psi=\nabla \varphi, \qquad \varphi\in H^2(M),$$
and consequently we obtain  the following decomposition for general symmetric $(0, 2)$ tensor fields:
\begin{equation}
u_{ij}=v_{ij}+ D_iD_j \varphi, \qquad D^iv_{ij}=0, \qquad \varphi \in H^2(M). \la{02d}
\end{equation}

Finally, for the symmetric and divergence free $(0, 2)$  field  $\delta F(g_{ij})$, which is the Euler-Lagrangian 
of the Einstein-Hilbert functional  in (\ref{el}) and is orthogonal to all divergence-free fields, there is a scalar field $\varphi \in H^2(M)$  such that 
$$\delta F(g_{ij}) = D_i D_j \varphi, $$
which, by adding the energy-momentum tensor $T_{ij}$, leads to the new gravitational field equations (\ref{1.3}). 

We remark here that the orthogonal decompositions (\ref{rsd})  and (\ref{02d}) are reminiscent of the orthogonal decomposition of vectors fields into gradient and divergence parts, which are crucial for studying 
incompressible  fluid flows; see among many others \cite{amsbook, ptd}.

This article is divided into two parts. The physically inclined readers can go directly to the physics part after reading this Introduction. 

\part{Mathematics}
\section{\large Preliminaries}
\subsection{Sobolev spaces of tensor fields}
Let $(M,g_{ij})$ be an $n$-dimensional Riemannian manifold with
metric $(g_{ij})$, and $E=T^r_sM$ be an $(r,s)$-tensor bundle on $M$.
A mapping  $u: M\to  E$
is called a section of the tensor-bundle $E$  or a tensor field. In
a local coordinate system  $x$, a tensor field $u$ can be expressed 
component-wise as follows:
$$
u=\left\{u^{j_1\cdots j_r}_{i_1\cdots i_s}(x)\quad |\quad  1\leq i_1,\cdots
,i_s, j_1,\cdots ,j_r\leq n\right\},
$$ 
where $u^{j_1\cdots j_r}_{i_1\cdots i_s}(x)$ are functions of $x\in U$. 
The section $u$ is called $C^r$-tensor
field or $C^r$-section  if its components are $C^r$-functions.

For any real number $1\leq p<\infty$, let $L^p(E)$ be the space of
all $L^p$-integrable sections of $E$:
$$L^p(E)=\left\{u: M\rightarrow E\quad \Big|\quad  \int_M|u|^pdx < \infty\right\},$$
equipped with the norm
$$||u||_{L^p}=\left[\int_M|u|^pdx\right]^{{1}/{p}}=\left[\int_M\sum|u^{j_1\cdots j_r}_{i_1\cdots
i_s}|^pdx\right]^{1/p}.
$$ 
For $p=2$, $L^2(E)$ is a Hilbert space equipped with the inner
product
\begin{equation}
(u,v)=\int_Mg_{j_1k_1}\cdots g_{j_rk_r}g^{i_1l_1}\cdots
g^{i_sl_s}u^{j_1\cdots j_r}_{i_1\cdots i_s}v^{k_1\cdots
k_r}_{l_1\cdots l_s}\sqrt{-g}dx,\label{(2.1)}
\end{equation}
where $(g_{ij})$ is Riemannian metric, $(g^{ij})=(g_{ij})^{-1}$,  $g={\rm det}(g_{ij})$, and 
$\sqrt{-g}dx$  is  the volume
element.

For any positive integer $k$ and any real number $1\le p < \infty$, we can also define the Sobolev spaces $W^{k,p}(E)$ to be the subspace of $L^p(E)$ such that all covariant derivatives of u up to order $k$ are in $L^p(E)$. The norm of 
$W^{k,p}(E)$  is always denoted by $\| \cdot \|_{W^{k, p}}$.

As $p=2$, the spaces $W^{k,p}(E)$ are Hilbert spaces, and are 
usually denoted by
\begin{equation}
H^k(E)=W^{k,2}(E)\qquad \text{ for }  k\geq 0, \label{(2.3)}
\end{equation}
equipped with inner product $(\cdot ,\cdot )_{H^k}$  and norm $\| \cdot \|_{H^k}$.

\subsection{Gradient and divergent operators}

Let $u: M\rightarrow E$ be an $(r,s)$-tensor field, with the local
expression
\begin{equation}
u=\left\{u^{j_1\cdots j_r}_{i_1\cdots i_s}\right\}.\label{(2.5)}
\end{equation}
Then the gradient of $u$ is defined as
\begin{equation}
\nabla u=\{D_ku^{j_1\cdots j_r}_{i_1\cdots i_s}\},\label{(2.6)}
\end{equation}
where $D=(D_1,\cdots ,D_n)$ is the covariant derivative. It is clear that
the gradient $\nabla u$ defined by (\ref{(2.6)}) is an
$(r,s+1)$-tensor field:
$$
\nabla u: M\rightarrow T^r_{s+1}M.
$$ 
We define $\nabla^\ast u$ as
\begin{equation}
\nabla^\ast u=\{g^{kl}D_lu\}:\ M\rightarrow T^{r+1}_sM\ \ \ \ {\rm
for}\ u\ \ \text{\rm as in (\ref{(2.5)})}.\label{(2.7)}
\end{equation}

For an $(r+1,s)$-tensor field
$u=\{u^{j_1\cdots l\cdots j_r}_{i_1\cdots i_s}\},$
the divergence of $u$ is defined by
\begin{equation}
\text{\rm div } u=\{D_lu^{j_1\cdots l\cdots j_r}_{i_1\cdots
i_s}\}.\label{(2.8)}
\end{equation}
Therefore, the divergence $\text{\rm div } u$ defined by (\ref{(2.8)}) is
an $(r,s)$-tensor field. Likewise, for an $(r,s+1)$-tensor field
$$u=\{u^{j_1\cdots j_r}_{i_1\cdots l\cdots i_s}\},$$
the following operator is also called the divergence of $u$,
\begin{equation}
\text{\rm div } u=\{D^lu^{j_1\cdots j_r}_{i_1\cdots l\cdots
i_s}\},\label{(2.9)}
\end{equation}
where $D^l=g^{lk}D_k$, which is an $(r,s)$-tensor field.

For the gradient operators (\ref{(2.6)})-(\ref{(2.7)}) and the
divergent operators (\ref{(2.8)})-(\ref{(2.9)}), it is well known
that the following integral formulas hold true; see among others \cite{CLN}.

\bt\la{t2.1}
Let $(M,g_{ij})$ be a closed Riemannian
manifold. If $u$ is an $(r-1,s)$-tensor and $v$ is an $(r,s)$
tensor, then we have
\begin{equation}
(\nabla^\ast u,v) =- (u,{\rm div}
v),\label{(2.10)}
\end{equation}
where $\nabla^\ast u$  is  as in (\ref{(2.7)}) and $\text{\rm div } v$  is  as in
(\ref{(2.8)}), the inner product $(\cdot ,\cdot )$ is as defined by
(\ref{(2.1)}). If $u$ is an $(r,s-1)$-tensor and $v$ is an $(r,s)$
tensor, then
\begin{equation}
(\nabla u,v) =- (u,{\rm
div}v),\label{(2.11)}
\end{equation}
where $\nabla u$  is  as in (\ref{(2.6)}) and ${\rm div} v$ is as in
(\ref{(2.9)}).

\et

\br
\la{r2.2}
{\rm
 If $M$ is a manifold with boundary $\partial
M\neq\emptyset$, and $u|_{\partial M}=0$ or $v|_{\partial M}=0$, then the
formulas (\ref{(2.10)}) and (\ref{(2.11)}) still hold true.
}
\er

\subsection{Acute-angle principle}

Let $H$ be a Hilbert space equipped with inner product $(\cdot, \cdot)$ and norm $\| \cdot \|$, and
$G:\ H\to H$  be a mapping. We say that $G$ is weakly continuous if for any sequence
$\{u_n\}\subset H$ weakly converging to $u_0$, i.e.
$$u_n\rightharpoonup u_0\ {\rm in}\ H,$$
we have
$$
\lim\limits_{n\rightarrow\infty} (Gu_n,v)= (Gu_0,v)\qquad  \forall
v\in H.
$$ 
If the operator $G$ is linear and
bounded, then $G$ is weakly continuous. 
The following theorem is called  the acute-angle principle \cite{MA-PDE}.

\bt\la{t2.3}
If a mapping $G:\ H\to H$
is weakly continuous, and satisfies 
$$ (Gu, u) \geq\alpha ||u||^2 -\beta ,$$
for some constants $\alpha ,\beta >0$, then for any $f\in H$ there
is a $u_0\in H$ such that
$$ (Gu_0-f,v) =0\qquad  \forall v\in H.$$
\et

\section{Orthogonal Decomposition for Tensor Fields} 
\subsection{Main theorems}

The aim of this section is to derive an  orthogonal
decomposition for  $(r,s)$-tensor fields with $r+s\geq 1$ into divergence-free and gradient parts. This decomposition plays a crucial  role for the theory  of gravitational
field, dark matter and dark energy developed later in this article.

Let $M$ be a closed Riemannian manifold, and $v\in
L^2(T^r_sM)$  $(r+s\geq 1)$. We say  that  $v$ is divergence-free, i.e.,  $\text{\rm
div } v=0$, if for all $\nabla\psi\in L^2(T^r_sM)$, 
\begin{equation}
(v,\nabla\psi ) =0.\label{(3.1)}
\end{equation}
Here $\psi\in H^1(T^{r-1}_sM)$ or $H^1(T^r_{s-1}M)$, and $(\cdot ,\cdot )$  is the $L^2$-inner 
product defined by 
(\ref{(2.1)}).

We remark that if $v\in H^1(T^r_sM)$ satisfies (\ref{(3.1)}), then
$v$ is weakly differentiable, and ${\rm div } v=0$ in the  $L^2$-sense.
If $v\in L^2(T^r_sM)$ is not differential, then (\ref{(3.1)}) means
that $v$ is divergence-free in the distribution sense.

\bt[Orthogonal Decomposition Theorem]  \la{t3.1} Let $M$ be
a closed Riemannian manifold, and $u\in L^2(T^r_sM)$ with $r+s\geq
1$. The following assertions  hold true:

\begin{itemize}
\item[(1)]  The tensor field $u$ has the following orthogonal decomposition:
\begin{equation}
u=\nabla\varphi +v,\label{(3.2)}
\end{equation}
where $\varphi\in H^1(T^{r-1}_sM)$ or $\varphi\in H^1(T^r_{s-1}M)$,
and  $\text{\rm div } v=0$.

\item[(2)] If $M$ is compact, then $u$ can be orthogonally decomposed into
\begin{equation}
u=\nabla\varphi +v+h,\label{(3.3)}
\end{equation}
where $\varphi$ and $v$ are as in (\ref{(3.2)}), and $h$ is a harmonic
field, i.e.
\begin{equation}
\text{ div } v=0,\ \ \ \ \text{\rm div } h=0,\ \ \ \ \nabla h=0.\label{(3.4)}
\end{equation}
In particular the subspace of all harmonic tensor fields in $L^2(T^r_sM)$
is of finite dimensional:
\begin{equation} 
H(T^r_sM)=\{h\in L^2(T^r_sM)|\ \nabla h=0,\ \text{\rm div }
h=0\}, \qquad \text{\rm dim } H<\infty. \label{(3.5)}
\end{equation}
\end{itemize}
\et

\br\la{r3.2}
{\rm
 The above orthogonal decomposition theorem  implies
that $L^2(E)$   $ (E=T^r_sM)$ can be decomposed into
\begin{align*}
&L^2(E)=G(E)\oplus L^2_D(E) && \text{\rm for}\ \partial M=\emptyset,\\
&L^2(E)=G(E)\oplus H(E)\oplus L^2_N(E)&& \text{\rm for}\ M\ {\rm
compact}. \end{align*} 
Here $H$ is as in (\ref{(3.5)}), and
\begin{align*}
&G(E)=\{v\in L^2(E)\quad |\quad  v=\nabla\varphi, \varphi\in H^1(T^r_{s-1}M)\},\\
&L^2_D(E)=\{v\in L^2(E)\quad |\quad  \text{\rm div}\ v=0\},\\
&L^2_N(E)=\{v\in L^2_D(E)\quad |\quad  \nabla v\neq 0\}.
\end{align*}
They are orthogonal to each other:
$$L^2_D(E)\bot G(E),\ \ \ \ L^2_N(E)\bot H(E),\ \ \ \ G(E)\bot
H(E).$$
}
\er

\br\la{r3.3}
{\rm 
 The dimension of the harmonic space $H(E)$ is
related with the bundle structure of $E=T^r_sM$. It is conjectured
that
$${\rm dim} H=k= \text {\rm the degree of freedom of }\ E.$$
Namely  $k$ is the integer that $E$ can be decomposed into the
Whitney sum of a $k$-dimensional trivial bundle $E^k=M\times \R^k$
and a nontrivial bundle $E_1$, i.e.
$$E=E_1\oplus E^k.$$
}
\er

\bp[Proof of Theorem \ref{t3.1}]
We proceed in several steps as follows.

\medskip

{\sc Step 1  Proof of  Assertion (1).} Let  $u\in L^2(E)$, $E=T^r_sM$ 
$(r+s\geq 1)$. Consider the equation
\begin{equation}
\Delta\varphi =\text{\rm div } u\ \ \ \ \text{\rm in } M,\label{(3.6)}
\end{equation}
where $\Delta$ is the Laplace operator defined by
\begin{equation}
\Delta  ={\rm div} \nabla .\label{(3.7)}
\end{equation}

Without loss of generality, we only consider the case where  $\text{\rm
div } u\in\tilde{E}=T^{r-1}_sM$. It is clear that if the equation
(\ref{(3.6)}) has a solution $\varphi\in H^1(\tilde{E})$, then by
(\ref{(3.7)}), the following vector field must be divergence-free
\begin{equation}
v=u-\nabla\varphi\in L^2(E). \label{(3.8)}
\end{equation}
Moreover, by (\ref{(3.1)}) we have
\begin{equation}
(v,\nabla\varphi ) =0. \label{(3.9)}
\end{equation}
Namely $v$ and $\nabla\varphi$ are  orthogonal. Therefore, the orthogonal decomposition
$u=v+\nabla\varphi$   follows from 
(\ref{(3.8)}) and (\ref{(3.9)}).

It suffices  then to prove that (\ref{(3.6)}) has a weak solution
$\varphi\in H^1(\tilde{E})$:
\begin{equation}
(\nabla\varphi -u,\nabla\psi ) =0\qquad 
\forall\psi\in H^1(\tilde{E}).\label{(3.10)}
\end{equation}
To this end, let 
\begin{align*}
& H=H^1(\tilde{E}) \setminus  \tilde{H},\\
&  \tilde{H}=\{\psi\in
H^1(\tilde{E})|\ \nabla\psi =0\}.
\end{align*}
 Then we define a linear  operator $G:
H\rightarrow H$ by
\begin{equation}
(G\varphi ,\psi ) =(\nabla\varphi ,\nabla\psi ) \qquad \forall\psi\in H.\label{(3.11)}
\end{equation}
It is clear that the linear operator $G: H\rightarrow H$ is bounded, weakly continuous, and
\begin{equation}
(G\varphi ,\varphi )=(\nabla\varphi ,\nabla\varphi
)=||\varphi||^2. \label{(3.12)}
\end{equation}
Based on Theorem \ref{t2.3}, for any $f\in H$,  the equation
$$\Delta\varphi =f\ \ \ \ {\rm in}\ M$$
has a weak solution $\varphi\in H$. Hence for $f={\rm div} u$ the
equation (\ref{(3.6)}) has a solution, and Assertion (1) is
proved. In fact the solution of (\ref{(3.6)}) is unique. We remark
that by the Poincar\'e  inequality, for the space
$H=H^1(\tilde{E})\setminus \{\psi |\nabla\psi =0\}$, (\ref{(3.12)}) is an
equivalent norm of $H$. In addition, by Theorem \ref{t2.1}, the weak formulation 
(\ref{(3.10)})  for (\ref{(3.6)}) is well-defined.

\medskip

{\sc Step 2 Proof of  Assertion (2).} 
Based on Assertion (1), we have 
\begin{eqnarray*}
&&H^k(E)=H^k_D\oplus G^k,\\
&&L^2(E)=L^2_D\oplus G,
\end{eqnarray*}
where
\begin{eqnarray*}
&&H^k_D=\{u\in H^k(E)\quad |\quad  \text{\rm div} u=0\}, \\
&&G^k=\{u\in H^k(E)\quad |\quad  u=\nabla\psi\}.
\end{eqnarray*}
Define an operator $\tilde{\Delta}: H^2_D(E)\rightarrow L^2_D(E)$ by
\begin{equation}
\tilde{\Delta}u=P\Delta u,\label{(3.13)}
\end{equation}
where $P: L^2(E)\rightarrow L^2_D(E)$ is the canonical orthogonal projection.

We know that  the  Laplace operator $\Delta$ can be expressed as
\begin{equation}
\Delta ={\rm div}  \nabla
=D^kD_k=g^{kl}\frac{\partial^2}{\partial x^k\partial
x^l}+B,\label{(3.14)}
\end{equation}
where $B$ is the lower order derivative linear operator. Since  $M$ is compact, 
 the Sobolev embeddings
$H^2(E)\hookrightarrow H^1(E)\hookrightarrow L^2(E)$ are compact,
which implies that the lower order derivative operator
$$B: H^2(M, \R^N)\rightarrow L^2(M, \R^N)\ \text{\rm is compact},$$
where the integer $N$ is the dimension of the  tensor bundle $E$. According to
the elliptic operator theory, the elliptic operator in
(\ref{(3.14)}) 
$$
A=g^{kl}\frac{\partial^2}{\partial x^k\partial x^l}:\
H^2(M,\R^N)\rightarrow L^2(M,\R^N)
$$ 
is a linear homeomorphism.
Therefore the operator in (\ref{(3.14)}) is a linear completely
continuous field
$$\Delta :\ H^2(E)\rightarrow L^2(E),$$
which implies  that the operator of (\ref{(3.13)}) is also a
linear completely continuous field:
$$\tilde{\Delta}=P\Delta :\ H^2_D(E)\rightarrow L^2_D(E).$$
 
 By the the spectral theorem of completely continuous fields \cite{b-book, ptd}, the space
$$\tilde{H}=\{u\in H^2_D(E)|\ \tilde{\Delta}u=0\}$$
is finite dimensional,   and  is the eigenspace of the  eigenvalue
$\lambda =0$. By Theorem \ref{t2.1}, for $u\in\tilde{H}$
\begin{align*}
\int_M(\tilde{\Delta}u,u)\sqrt{-g}dx=&\int_M(\Delta u,u)\sqrt{-g}dx &&  (\text{\rm by div} u=0)\\
=&-\int_M(\nabla u,\nabla u)\sqrt{-g}dx\\
=&0 && (\text{\rm by} \tilde{\Delta}u=0).
\end{align*}
It follows that
$$u\in\tilde{H}\Leftrightarrow \nabla u=0\ \Rightarrow
H=\tilde{H},$$ 
where $H$ is the harmonic space as in (\ref{(3.5)}).
Thus we have \begin{eqnarray*} 
&&L^2_D(E)=H\oplus L^2_N(E),\\
&&L^2_N(E)=\{u\in L^2_D(E)|\ \nabla u\neq 0\}. 
\end{eqnarray*} 
The proof of  Theorem \ref{t3.1} is complete.
\ep

\subsection{Uniqueness of the orthogonal decomposition}

In Theorem \ref{t3.1}, a tensor field $u\in L^2(T^r_sM)$ with $r+s\geq 1$
can be orthogonally decomposed into
\begin{equation}
\begin{aligned} 
& u=\nabla\varphi +v &&\text{\rm for}\ \partial
M=\emptyset ,\\
& u=\nabla\varphi +v+h&&\text{\rm for}\ M\ {\rm compact}.
\end{aligned}\label{(3.15)}
\end{equation}

Now we  address the uniqueness problem of the decomposition 
(\ref{(3.15)}). In fact, if $u$ is a vector field or a co-vector
field, i.e.
$$u\in L^2(TM)\ {\rm or}\ u\in L^2(T^*M),$$
then the decomposition of (\ref{(3.15)}) is unique.

We can see that if $u\in L^2(T^r_sM)$ with $r+s\geq 2$, then there
are different types of the decompositions of (\ref{(3.15)}). For
example, for $u\in L^2(T^0_2M)$, the local expression of $u$ is
given by
$$u=\{u_{ij}(x)\}.$$
In this case, $u$ has two types of decompositions:
\begin{eqnarray}
&&u_{ij}=D_i\varphi_j+v_{ij},\ \ \ \ D^iv_{ij}=0,\label{(3.16)}\\
&&u_{ij}=D_j\psi_i+w_{ij},\ \ \ \ D^iw_{ij}=0.\label{(3.17)}
\end{eqnarray}
It is easy to see that if $u_{ij}\neq u_{ji}$ then both
(\ref{(3.16)}) and (\ref{(3.17)}) are two different decompositions of
$u_{ij}$. Namely
$$\{v_{ij}\}\neq\{w_{ij}\},\ \ \ \ (\varphi_1,\cdots ,\varphi_n)\neq
(\psi_1,\cdots ,\psi_n).$$

If $u_{ij}=u_{ji}$ is symmetric, $u$ can be orthogonally decomposed
into the following two forms:
\begin{eqnarray*}
&&u_{ij}=v_{ij}+D_i\varphi_j,\ \ \ \ D^iv_{ij}=0,\\
&&u_{ij}=w_{ij}+D_j\psi_i,\ \ \ \ D^iw_{ij}=0,
\end{eqnarray*}
and $\varphi$ and $\psi$ satisfy
\begin{eqnarray}
&&\Delta\varphi_j=D^ku_{kj},\label{(3.18)}\\
&&\Delta\psi_j=D^ku_{jk}.\label{(3.19)}
\end{eqnarray}
By $u_{ij}=u_{ji}$ we have $D^ku_{kj}=D^ku_{jk}$. Hence,
(\ref{(3.18)}) and (\ref{(3.19)}) are the same, and $\varphi =\psi$. 
Therefore, the symmetric tensors $u_{ij}$ can be
written as
\begin{eqnarray}
&&u_{ij}=v_{ij}+D_i\varphi_j,\ \ \ \ D^iv_{ij}=0,\label{(3.20)}\\
&&u_{ij}=w_{ij}+D_j\varphi_i,\ \ \ \ D^jw_{ij}=0.\label{(3.21)}
\end{eqnarray}
From (\ref{(3.20)})-(\ref{(3.21)}) we can deduce the following
theorem.

\bt\la{t3.4}
 Let $u\in L^2(T^0_2M)$ be symmetric, i.e.
$u_{ij}=u_{ji}$, and the first Betti number $\beta_1(M)=0$ for $M$.
Then the following assertions hold true:

\begin{itemize}
\item[(1)] $u$ has a  unique orthogonal decomposition if and only if there is
a scalar function $\varphi\in H^2(M)$ such that $u$ can be expressed as
\begin{equation}
\begin{array}{l}
u_{ij}=v_{ij}+D_iD_j\varphi ,\\
v_{ij}=v_{ji},\ \ \ \ D^iv_{ij}=0.
\end{array}\label{(3.22)}
\end{equation}

\item[(2)] $u$ can be orthogonally decomposed in the form of (\ref{(3.22)})
if and only if $u_{ij}$ satisfy
\begin{equation}
\frac{\partial}{\partial x^j}(D^ku_{ki})-\frac{\partial}{\partial
x^i}(D^ku_{kj})\label{(3.23)}= \frac{\partial}{\partial
x^i}\left(R^k_j\frac{\partial\varphi}{\partial
x^k}\right)-\frac{\partial}{\partial
x^j}\left(R^k_i\frac{\partial\varphi}{\partial x^k}\right),
\end{equation}
where $R^k_j=g^{ki}R_{ij}$   and   $R_{ij}$ are the Ricci curvature tensors.

\item[(3)] If $v_{ij}$ in (\ref{(3.20)}) is symmetric: $v_{ij}=v_{ji}$,  then
$u$ can be expressed by (\ref{(3.22)}).
\end{itemize}
\et

\bp 
We only need to prove Assertions (2) and (3).

We first prove Assertion (2). It follows from (\ref{(3.20)})
that
\begin{equation}
\frac{\partial}{\partial x^j}(D^ku_{ki})-\frac{\partial}{\partial
x^i}(D^ku_{kj})=\frac{\partial\Delta\varphi_i}{\partial
x^j}-\frac{\partial\Delta\varphi_j}{\partial x^i},\label{(3.24)}
\end{equation}
where $\Delta =D^kD_k$. 
By the Weitzenb\"ock formula \cite{MA-topology},
\begin{equation}
\Delta\varphi_i=-(\delta d+d\delta
)\varphi_i-R^k_i\varphi_k,\label{(3.25)}
\end{equation}
and $(\delta d+d\delta )$ is the Laplace-Beltrami operator. We know
that for $\omega =\varphi_idx^i$, 
\begin{align*} 
&d\omega =0 && 
\Leftrightarrow  && 
\varphi_i=\frac{\partial\varphi}{\partial x^i}, \\
&d\delta\omega =\nabla (\tilde{\Delta}\varphi ) &&  \Leftrightarrow &&
\varphi_i=\frac{\partial\varphi}{\partial x^i},
\end{align*}
where $\nabla$ is the gradient operator, and
$$\tilde{\Delta}\varphi =-\frac{1}{\sqrt{-g}}\frac{\partial}{\partial
x^i}\left(\sqrt{-g}g^{ij}\frac{\partial\varphi}{\partial
x^j}\right).$$ Namely $$(\delta d+d\delta
)\varphi_i=\frac{\partial}{\partial x^i}\tilde{\Delta}\varphi\
\Leftrightarrow\ \varphi_i=\frac{\partial\varphi}{\partial x^i}.$$
Hence, we infer from (\ref{(3.25)}) that
\begin{equation}
\Delta\varphi_i=-\frac{\partial}{\partial x^i}\tilde{\Delta}\varphi
-R^k_i\frac{\partial\varphi}{\partial x^k}\ \Leftrightarrow\
\varphi_i=\frac{\partial\varphi}{\partial x^i}.\label{(3.26)}
\end{equation}
Thus, by (\ref{(3.24)}) and (\ref{(3.26)}),  we obtain  that
$$\frac{\partial}{\partial x^j}(D^ku_{ki})-\frac{\partial}{\partial
x^i}(D^ku_{kj})=\frac{\partial}{\partial
x^i}(R^k_j\frac{\partial\varphi}{\partial
x^k})-\frac{\partial}{\partial
x^j}(R^k_i\frac{\partial\varphi}{\partial x^k})$$ holds true if and
only if the tensor $\psi =(\varphi_1,\cdots ,\varphi_n)$ in
(\ref{(3.20)}) is a gradient $\psi =\nabla\varphi$. Assertion
(2) is proven.

Now we verify Assertion (3). Since $v_{ij}$ in (\ref{(3.20)}) is
symmetric, we have 
\begin{equation}
D_i\varphi_j=D_j\varphi_i.\label{(3.27)}
\end{equation}
Note that
\begin{equation}
D_i\varphi_j=\frac{\partial\varphi_j}{\partial
x^i}-\Gamma^k_{ij}\varphi_k,\label{(3.28)}
\end{equation}
where $\Gamma^k_{ij}$ is the Levi-Civita connection, and
$\Gamma^k_{ij}=\Gamma^k_{ji}$. We infer then from (\ref{(3.27)}) that
\begin{equation}
\frac{\partial\varphi_j}{\partial
x^i}=\frac{\partial\varphi_i}{\partial x^j}.\label{(3.29)}
\end{equation}
By assumption,  the 1-dimensional homology of $M$ is zero,
$$H_1(M)=0,$$
and it follows from the de Rham theorem and (\ref{(3.29)}) that 
$$
\varphi_k=\frac{\partial\varphi}{\partial x^k}, 
$$ 
for some scalar
function  $\varphi$.  
Thus Assertion (3) follows and  the proof
is complete. 
\ep

\br 
{\rm
The conclusions of Theorem \ref{t3.4} are also valid
for second-order contra-variant symmetric tensors $u=\{u^{ij}\}$, and the decomposition is given as follows:
\begin{align*}
& u^{ij}=v^{ij}+g^{ik}g^{jl}D_kD_l\varphi ,\\
& D_iv^{ij}=0,\ v^{ij}=v^{ji},\ \varphi\in H^2(M).
\end{align*}
}
\er

\section{Variational Principle for Functionals of Riemannian
Metric} \setcounter{equation}{0}
\subsection{General theory}

Hereafter we always assume that $M$ is a closed manifold. A
Riemannian metric $G$ on $M$ is a mapping
\begin{equation}
G:\ M\rightarrow T^0_2M=T^*M\otimes T^*M,\label{(4.1)}
\end{equation}
which is symmetric and nondegenerate, i.e., in a local coordinate
$(v,x), G$ can be expressed as
\begin{equation}
G=\{g_{ij}(x)\}\ \ \ \ {\rm with}\ \ \ \ g_{ij}=g_{ji},\label{(4.2)}
\end{equation}
and the matrix $(g_{ij})$ is invertible on $M$:
\begin{equation}
(g^{ij})=(g_{ij})^{-1}.\label{(4.3)}
\end{equation}

If we regard a Riemannian metric $G=\{g_{ij}\}$ as a tensor field on
manifold $M$, then the set of all metrics $G=\{g_{ij}\}$ on $M$
constitutes a topological space, called the space of Riemannian
metrics on $M$. We denote
\begin{equation}
G^{-1}=\{g^{ij}\}:\ M\rightarrow T^2_0M=TM\otimes TM.\label{(4.4)}
\end{equation}
The space for  Riemannian metrics on $M$ is denied by 
\begin{align}
W^{m,2}(M,g)= &  \Big\{G \ |\ G\in W^{m,2}(T^0_2M),G^{-1}\in W^{m,2}(T^2_0M),\label{(4.5)}\\
 &  G  \text{ is the Riemannian metric on  $M$  as in (\ref{(4.2)})}\},\nonumber 
\end{align}
which is a metric space, but not a Banach space. However, it is a subspace of the direct sum  
of two Sobolev  spaces $W^{m,2}(T^0_2M)$ and $W^{m,2}(T^2_0M)$: 
$$
W^{m,2}(M,g)\subset W^{m,2}(T^0_2M)\oplus W^{m,2}(T^2_0M).
$$

A functional defined on $W^{m,2}(M,g):$
\begin{equation}
F:\ W^{m,2}(M,g)\rightarrow \R \label{(4.6)}
\end{equation}
is called the functional of Riemannian metric. Usually, the
functional (\ref{(4.6)}) can be expressed as
\begin{equation}
F(g_{ij})=\int_Mf(x,g_{ij},\cdots
,\partial^mg_{ij})\sqrt{-g}dx. \label{(4.7)}
\end{equation}

Since $(g^{ij})$ is the inverse of $(g_{ij})$, we have
\begin{equation}
g_{ij}=\frac{1}{g}\times \text{\rm the cofactor of}\ g^{ij}.\label{(4.8)}
\end{equation}
Therefore, $F(g_{ij})$ in (\ref{(4.7)}) also depends on $g^{ij}$,
i.e. putting (\ref{(4.8)}) in (\ref{(4.7)}) we get
\begin{equation}
F(g^{ij})=\int_M\tilde{f}(x,g^{ij},\cdots
,D^mg_{ij})\sqrt{-g}dx.\label{(4.9)}
\end{equation}

We note that although $W^{m,2}(M,g)$ is not a linear space, for a
given element $g_{ij}\in W^{m,p}(M,g)$ and any symmetric tensor
$X_{ij}$ and $X^{ij}$, there is a number $\lambda_0>0$ such that
\begin{equation}
\begin{aligned}
& g_{ij}+\lambda X_{ij}\in W^{m,2}(M,g) &&  \forall 0\leq |\lambda
|<\lambda_0,\\
& g^{ij}+\lambda X^{ij}\in W^{m,2}(M,g)  &&  \forall 0\leq |\lambda
| <\lambda_0.
\end{aligned}\label{(4.10)}
\end{equation}

Due to (\ref{(4.10)}), we can define the derivative operators of
the functional $F$, which are also called the Euler-Lagrange operators of
$F$, as follows
\begin{eqnarray}
&&\delta_*F:\ W^{m,2}(M,g)\rightarrow
W^{-m,2}(T^2_0M),\label{(4.11)}\\
&&\delta^*F:\ W^{m,2}(M,g)\rightarrow
W^{-m,2}(T^0_2M),\label{(4.12)}
\end{eqnarray}
where $W^{-m,2}(E)$ is the dual space of $W^{m,2}(E)$, and
$\delta_*F, \delta^*F$ are given by
\begin{eqnarray}
&& (\delta_*F(g_{ij}),X) =\frac{d}{d\lambda}F(g_{ij}+\lambda
X_{ij})|_{\lambda =0},\label{(4.13)}\\
&&(\delta^*F(g^{ij}),X) =\frac{d}{d\lambda}F(g^{ij}+\lambda
X^{ij})|_{\lambda =0}.\label{(4.14)}
\end{eqnarray}

For any given metric $g_{ij}\in W^{m,2}(M, g)$, the value of $\delta_*F$ and
$\delta^*F$ at $g_{ij}$ are second-order contra-variant and covariant
tensor fields respectively, i.e.
\begin{eqnarray*}
&&\delta_*F(g_{ij}):\ M\rightarrow TM\otimes TM,\\
&&\delta^*F(g_{ij}): M\rightarrow T^*M\otimes T^*M.
\end{eqnarray*}
Moreover, the equations
\begin{eqnarray}
&&\delta_*F(g_{ij})=0,\label{(4.15)}\\
&&\delta^*F(g_{ij})=0,\label{(4.16)}
\end{eqnarray}
are called the Euler-Lagrange equations of $F$, and the solutions of
(\ref{(4.15)}) and (\ref{(4.16)}) are called the extremum points or
critical points of $F$.

\bt\la{t4.1}
 Let $F$ be the functionals defined by
(\ref{(4.6)}) and (\ref{(4.9)}). Then the following assertions hold true:

\begin{itemize}

\item[(1)] For any $g_{ij}\in W^{m,2}(M,g), \delta_*F(g_{ij})$ and
$\delta^*F(g_{ij})$ are symmetric tensor fields.

\item[(2)] If $\{g_{ij}\}\in W^{m,2}(M,g)$ is the extremum point of $F$,
then $\{g^{ij}\}$ is also an extremum point of $F$, i.e.
$\{g_{ij}\}$ satisfies (\ref{(4.15)}) and (\ref{(4.16)})
if and only if   $\{g^{ij}\}$ satisfies (\ref{(4.15)}) and
(\ref{(4.16)}).

\item[(3)] $\delta_*F$ and $\delta^*F$ have the following relation
$$(\delta^*F(g_{ij}))^{kl}=-g^{kr}g^{ls}(\delta_*F(g_{ij}))_{rs},$$
where $(\delta^*F)^{kl}$ and $(\delta_*F)_{kl}$ are the components
of $\delta^*F$ and $\delta_*F$ respectively.
\end{itemize}
\et

\bp 
We only need to verify  Assertion (3). Noting that
$$g_{ik}g^{kj}=\delta^j_i,$$
we have the variational relation
$$\delta (g_{ik}g^{kj})=g_{ik}\delta g^{kj}+g^{kj}\delta g_{ik}=0.$$
It implies that
\begin{equation}
\delta g^{kl}=-g^{ki}g^{lj}\delta g_{ij}.\label{(4.17)}
\end{equation}
In addition, in (\ref{(4.13)}) and (\ref{(4.14)}),
$$\lambda X_{ij}=\delta g_{ij},\ \ \ \ \lambda X^{ij}=\delta
g^{ij},\\ \ \ \lambda\neq 0\ {\rm small}.$$ Therefore, by
(\ref{(4.17)}) we get
$$
((\delta_*F)_{kl},\delta
g^{kl}) =- ((\delta_*F)_{kl},g^{ki}g^{lj}\delta g_{ij}) =(-g^{ki}g^{lj}(\delta_*F)_{kl},\delta g_{ij}) =
 ((\delta^*F)^{ij},\delta g_{ij}).
$$
Hence 
$$(\delta^*F)^{ij}=-g^{ki}g^{lj}(\delta_*F)_{kl}.$$
Thus Assertion (3) follows and the proof is complete.
\ep

\subsection{Scalar potential theorem for constraint variations}

We know that the critical  points of the functional $F$ in (\ref{(4.6)})
are the solution
\begin{equation}
\delta F(g_{ij})=0,\label{(4.18)}
\end{equation}
in the following sense
\begin{align}
(\delta F(g_{ij}), X) =&\frac{d}{d\lambda}F(g^{ij}+\lambda
X^{ij})|_{\lambda =0}\label{(4.19)}\\
=&\int_M(\delta F(g_{ij}))_{kl}X^{kl}\sqrt{-g}dx\nonumber\\
=&0\qquad  \forall X^{kl}=X^{lk}\ {\rm in}\ L^2(E),\nonumber
\end{align}
where $E=TM\otimes TM$. Hence, the critical  points of functionals of
Riemannian metrics are not solutions of (\ref{(4.18)}) in the usual
sense.

It is easy to see that $L^2(TM\otimes TM)$ can be orthogonally
decomposed into the direct sum of the symmetric and contra-symmetric
spaces, i.e.
\begin{eqnarray}
&&L^2(E)=L^2_s(E)\oplus L^2_c(E),\nonumber\\
&&L^2_s(E)=\{u\in L^2(E)|\ u_{ij}=u_{ji}\},\label{(4.20)}\\
&&L^2_c(E)=\{v\in L^2(E)|\ v_{ij}=-v_{ij}\}.\nonumber
\end{eqnarray}
Since $\delta F$ is symmetric, by (\ref{(4.20)}) the extremum points
$\{g_{ij}\}$ of $F$ satisfy more general equality
\begin{equation}
(\delta F(g_{ij}), X)=0\qquad \forall X=\{X_{ij}\} \in
L^2(E).\label{(4.21)}
\end{equation}

Thus, we can say that the extremum points of functionals of
the Riemannian metrics are solutions of (\ref{(4.18)}) in the usual
sense of (\ref{(4.21)}), or are zero points of the variational
operators
$$\delta F:\ W^{m,2}(M,g)\rightarrow W^{-m,2}(E).$$

Now we consider the variations of $F$ under the divergence-free
constraint. In this case, the Euler-Lagrangian equations with symmetric divergence-free constraints 
are equivalent to the Euler-Lagrangian equations with general divergence-free constraints. Hence we have the following definition.

\bd\la{d4.2}
Let $F: W^{m,2}(M,g)\rightarrow \R$ be a
functional of Riemannian metric. A metric tensor $\{g_{ij}\}\in
W^{m,2}(M,g)$ is called  an extremum point of $F$ with divergence-free
constraint, if $\{g_{ij}\}$ satisfies
\begin{equation}
(\delta F(g_{ij}),X) =0\qquad  \forall X=\{X_{ij}\}\subset
L^2_D(E),\label{(4.22)}
\end{equation}
where $L^2_D(E)$ is the space of all divergence-free tensors:
$$L^2_D(E)=\{X\in L^2(E)\quad |\quad  {\rm div}\ X=0\}.$$
\ed

It is clear that an extremum point satisfying (\ref{(4.22)}) is not a
solution of (\ref{(4.18)}). Instead, we have the scalar
potential theorem for the extremum points of divergence free
constraint (\ref{(4.22)}), which is based on the orthogonal
decomposition theorems. This result is also crucial for the 
gravitational field equations and the theory of dark matter and
dark energy developed later.

\bt[Scalar Potential Theorem]  \la{t4.3}
Assume that the first
Betti number of $M$ is zero, i.e. $\beta_1(M)=0$. Let $F$ be a 
functional of the Riemannian metric. Then there is a $\varphi\in
H^2(M)$ such that the extremum points $\{g_{ij}\}$ of $F$ with
divergence-free constraint  satisfy 
\begin{equation}
(\delta F(g_{ij}))_{kl}=D_kD_l\varphi .\label{(4.23)}
\end{equation}
\et

\bp
 Let $\{g_{ij}\}$ be an extremum point of $F$ under
the constraint  (\ref{(4.22)}). Namely, $\delta F(g_{ij})$ satisfies
\begin{equation}
\int_M(\delta F(g_{ij}))_{kl}X^{kl}\sqrt{-g}dx=0\qquad \forall
X=\{X_{kl}\}  \text{ with } D_kX^{kl}=0.\label{(4.24)}
\end{equation}
By Theorem \ref{t3.1}, $\delta_*F(g_{ij})$ can be orthogonally  decomposed as
\begin{equation}
(\delta F(g_{ij}))_{kl}=v_{kl}+D_k\psi_l,\ \ \ \
D^kv_{kl}=0.\label{(4.25)}
\end{equation}
By Theorem \ref{t2.1},  for any $D_kX^{kl}=0$, 
\begin{equation}
(D_k\psi_l,X^{kl})
=\int_MD_k\psi_lX^{kl}\sqrt{-g}dx
=  -\int_M\psi_lD_kX^{kl}\sqrt{-g}dx=0.\label{(4.26)}
\end{equation}
Therefore it follows from (\ref{(4.24)})-(\ref{(4.26)}) that
\begin{equation}
\int_Mv_{kl}X^{kl}\sqrt{-g}dx=0\qquad \forall
D_kX^{kl}=0.\label{(4.27)}
\end{equation}
Let $X^{kl}=g^{ki}g^{lj}v_{ij}$. Since 
$$D_kg_{ij}=D_kg^{ij}=0,$$
we have
$$
D_kX^{kl}= D_k(g^{ki}g^{lj}v_{ij})
=g^{lj}(g^{ik}D_kv_{ij})= g^{lj}D^iv_{ij}=0,
$$
thanks to  $D^iv_{ij}=0$.
Inserting $X^{kl}=g^{ki}g^{lj}v_{ij}$ into (\ref{(4.27)}) leads to
$$||v||_{L^2}^2=\int_Mg^{ki}g^{lj}v_{kl}v_{ij}\sqrt{-g}dx=0,$$
which implies that $v=0$. Thus, (\ref{(4.25)}) becomes
\begin{equation}
(\delta F(g_{ij}))_{kl}=D_k\psi_l.\label{(4.28)}
\end{equation}
By Theorem~\ref{t4.1}, $\delta F$ is symmetric. Hence we have
$$D_k\psi_l=D_l\psi_k.$$
It follows from (\ref{(3.28)}) that
\begin{equation}
\frac{\partial\psi_l}{\partial x^k}=\frac{\partial\psi_k}{\partial
x^l}.\label{(4.29)} 
\end{equation} 
By assumption, the first Betti
number of $M$ is zero, i.e. the 1-dimensional homology of $M$ is
zero: $H_1(M)=0$. It follows from the de Rham theorem that if
$$d(\psi_kdx^k)=\left(\frac{\partial\psi_k}{\partial
x^l}-\frac{\partial\psi_l}{\partial x^k}\right)dx^l\wedge dx^k=0,$$
then there  exists a scalar function $\varphi$ such that
$$d\varphi =\frac{\partial\varphi}{\partial x^k}dx^k=\psi_kdx^k.$$
Thus, we infer from (\ref{(4.29)}) that
$$
\psi_l=\frac{\partial\varphi}{\partial x^l}\ \text{\rm for some}\
\varphi\in H^2(M).
$$ 
Therefore we get (\ref{(4.23)}) from
(\ref{(4.28)}). The theorem is proved.
\ep

If the first Betti number $\beta_1(M)\neq 0$, then there are
$N=\beta_1(M)$ number of 1-forms:
\begin{equation}
\omega_{j}=\psi^{j}_kdx^k\in H^1_d(M) \quad   \text{  for } \quad 1\leq j\leq
N, \label{(4.30)}
\end{equation}
which constitute a basis of  the  1-dimensional de Rham homology
$H^1_d(M)$. We know that the components of $\omega_j$ are co-vector
fields:
\begin{equation}
\psi^{j}=(\psi^{j}_1,\cdots ,\psi^{j}_n)\in H^1(T^*M)\quad   \text{  for } \quad 1\leq j\leq
N, \label{(4.31)}
\end{equation}
which possess the following  properties:
$$\frac{\partial\psi^{j}_k}{\partial
x^l}=\frac{\partial\psi^{j}_l}{\partial x^k}\quad   \text{  for } \quad 1\leq j\leq
N,
$$ 
or equivalently,
$$D_l\psi^j_k=D_k\psi^j_l\quad   \text{  for } \quad 1\leq j\leq
N.$$
Namely, $\nabla\psi^{j}\in L^2(T^*M\otimes T^*M)$ are symmetric
second-order contra-variant tensors. Hence 
Theorem \ref{t4.3} can be extended to the non-vanishing first Betti number case as follows.

\bt\la{t4.4}
 Let the first Betti number $\beta_1(M)\neq 0$
for $M$. Then for the functional $F$ of Riemannian metrics, the
extremum points $\{g_{ij}\}$ of $F$ with the constraint (\ref{(4.22)})
satisfy the equations
\begin{equation}
(\delta F(g_{ij}))_{kl}=D_kD_l\varphi +\sum\limits^N_{j
=1}\alpha_{j}D_k\psi^{j}_l,\label{(4.32)}
\end{equation}
where $N=\beta_1(M), \alpha_j$ are  constants, $\varphi\in H^2(M)$,
and the tensors $\psi^{j}=(\psi^{j}_1,\cdots ,\psi^{j}_n)\in
H^1(T^*M)$ are as given by (\ref{(4.31)}).
\et

The proof of Theorem \ref{t4.4} is similar to Theorem \ref{t4.3}, and is omitted here.
\br\la{r4.5}
{\rm By the Hodge decomposition theory, the 1-forms
$\omega_{j}$ in (\ref{(4.30)}) are  harmonic:
$$d\omega_{j}=0,\ \ \ \ \delta\omega_{j}=0\quad   \text{  for } \quad 1\leq j\leq
N, $$ 
which implies that the tensors $\psi^{ j }$ in (\ref{(4.32)})
satisfy
\begin{equation}
(\delta d+d\delta )\psi^{ j }=0\quad   \text{  for } \quad 1\leq j\leq
N. \label{(4.33)}
\end{equation}
According to the Weitzenb\"ock formula (\ref{(3.25)}),
we obtain from (\ref{(4.33)})  that
\begin{equation}
D^kD_k\psi^{ j }_l=-R^k_l\psi^{ j }_k\quad   \text{  for } \quad 1\leq j\leq
N,\label{(4.34)}
\end{equation}
for $\psi^{ j }=(\psi^{ j }_1,\cdots ,\psi^{ j }_n)$ in
(\ref{(4.32)}).
}
\er

\br\la{r4.6}{\rm 
Theorem~\ref{t4.3} is derived for deriving  new gravitational field equation in the next section 
for explaining
the phenomena of dark matter and dark energy. The condition that
$\beta_1(M)=0$ means  that any loops in  the manifold  $M$ can shrink to a
point. Obviously, our universe  can be considered as a 4-dimensional manifold satisfying 
this condition.
}
\er

\part{Physics}
\section{Gravitational Field Equations}
\subsection{Einstein-Hilbert functional}
The general theory of relativity is based on three basic principles:
the principle of equivalence, the principle of general relativity, and the principle of Lagrangian dynamics. 
The first two principles tell us
that the spatial and temporal world is a 4-dimensional Riemannian
manifold $(M,g_{ij})$,  where the metric $\{g_{ij}\}$ represents
gravitational potential, and the third principle determines that  the Riemannian metric 
$\{g_{ij}\}$ is an extremum point of  the Lagrangian action, which is the
Einstein-Hilbert functional.

Let $(M,g_{ij})$ be an $n$-dimensional Riemannian manifold. The
Einstein-Hilbert functional\footnote{The matter tensor is included here as well.} 
\begin{equation} F: 
W^{2,2}(M, g)\rightarrow \R\label{(5.1)}
\end{equation}
is defined by
\begin{equation}
F(g_{ij})=\int_M\left(R+\frac{8\pi
G}{c^4}g^{ij}S_{ij}\right)\sqrt{-g}dx,\label{(5.2)}
\end{equation}
where $W^{2,2}(M, g)$ is  defined by (\ref{(4.5)}), $R=g^{kl}R_{kl}$
and $R_{kl}$ are the scalar and the Ricci curvatures, $S_{ij}$  is the stress
tensor, $G$ is the gravitational constant, and $c$ is the speed of light. 

The Euler-Lagrangian of  the  Einstein-Hilbert functional
$F$ is given by
\begin{equation}
\delta F(g_{ij})=R_{ij}- \frac{1}{2}g_{ij}R+\frac{8\pi
G}{c^4}T_{ij},\label{(5.3)}
\end{equation}
where $T_{ij}$ is the energy-momentum tensor given by
$$T_{ij}=S_{ij}-\frac{1}{2}g_{ij}S+g^{kl}\frac{\partial
S_{kl}}{\partial g^{ij}},\ \ \ \ S=g^{kl}S_{kl},
$$ 
and the Ricci  curvature 
tensor $R_{ij}$ is given by
\begin{eqnarray}
R_{ij}&=&\frac{1}{2}g^{kl}\left(\frac{\partial^2g_{kl}}{\partial
x^i\partial x^j}+\frac{\partial^2g_{ij}}{\partial x^k\partial
x^l}-\frac{\partial^2g_{il}}{\partial x^j\partial
x^k}-\frac{\partial^2g_{kj}}{\partial x^i\partial
x^l}\right)\label{(5.4)}\\
&&+g^{kl}g_{rs}\left(\Gamma^r_{kl}\Gamma^s_{ij}-\Gamma^r_{il}\Gamma^s_{kj}\right),\nonumber\\
\Gamma^k_{ij}&=&\frac{1}{2}g^{kl}\left(\frac{\partial
g_{il}}{\partial x^j}+\frac{\partial g_{jl}}{\partial
x^i}-\frac{\partial g_{ij}}{\partial x^l}\right).\la{(5.5)}
\end{eqnarray}
By (\ref{(5.3)})-(\ref{(5.5)}),  the Euler-Lagrangian  $\delta F(g_{ij})$ of 
the  Einstein-Hilbert functional is a
second order differential operator on $\{g_{ij}\}$, and $\delta
F(g_{ij})$ is symmetric.

\subsection{Einstein field equations}

The General Theory of Relativity consists  of two
main conclusions:

\begin{enumerate}
\item[1)] The space-time of our world is a 4-dimensional Riemannian manifold
$(M^4, g_{ij})$, and the metric $\{g_{ij}\}$ represents
gravitational potential.

\item[2)] The metric $\{g_{ij}\}$ is the extremum point of the
Einstein-Hilbert functional (\ref{(5.2)}). In other words,
gravitational field theory obeys the principle of Lagrange dynamics.

\end{enumerate}

The principle of
Lagrange dynamics is a universal principle, stated as:
 
 \medskip

\noindent {\bf Principle of Lagrange Dynamics. }
{\it For any physical system,
there are a set of state functions
$$u=(u_1,\cdots ,u_N),$$
which describe the state of this system, and there exists a
functional $L$ of $u$, called the Lagrange action:
\begin{equation}
L(u)=\int^T_0\int_{\Omega}\mathfrak{L}(u,Du,\cdots
,D^mu)dxdt,\label{(5.6)}
\end{equation}
such that the state $u$ is an extremum point of $L$. Usually the
function $\mathfrak{L}$ in (\ref{(5.6)}) is called the Lagrangian
density.
}

\medskip

Based on this principle, the gravitational field
equations are the Euler-Lagrange equations of the Einstein-Hilbert
functional:
\begin{equation}
\delta F(g_{ij})=0, \label{(5.7)}
\end{equation}
which are the classical Einstein field equations: 
\begin{equation}
R_{ij}-\frac{1}{2}g_{ij}R=-\frac{8\pi G}{c^4}T_{ij}.\label{(5.8)}
\end{equation}

By the Bianchi identities, the left hand side of (\ref{(5.8)}) is
divergence-free, i.e.
\begin{equation}
D^i(R_{ij}-\frac{1}{2}g_{ij}R)=0.\label{(5.9)}
\end{equation}
Therefore it is required in the general theory of relativity that the energy-momentum tensor $\{T_{ij}\}$ in (\ref{(5.8)})
satisfies  the following energy-momentum conservation law:
\begin{equation}
D^iT_{ij}=0\qquad \text{ for } 1\leq j\leq n.\label{(5.10)}
\end{equation}

\subsection{New gravitational field equations}
Motivated by the mystery of dark energy and dark matter and the other difficulties encountered by the Einstein field equations as mentioned in Introduction, we introduce in this section a new set of field equations, still obeying   the  three basic principles of the General Theory of Relativity. 

Our key observation  is a well-known fact that  the Riemannian metric $g_{ij}$  is divergence-free. This suggests us two important postulates for deriving  the new  gravitational field equations:

\begin{itemize}
\item The energy-momentum tensor of matter need not to be divergence-free  due to the presence of dark energy and dark matter; and 

\item The field equation obeys the Euler-Lagrange equation of the Einstein-Hilbert functional under the natural divergence-free constraint. 
\end{itemize}
Under these two postulates, by the Scalar Potential Theorem, Theorem~\ref{t4.3},  if  
the Riemannian metric $\{g_{ij}\}$ is an extremum
point of  the Einstein-Hilbert functional (\ref{(5.2)}) with the
divergence-free constraint (\ref{(4.22)}), then the
gravitational field equations are taken in the following form:
\begin{equation}
R_{ij}-\frac{1}{2}g_{ij}R=-\frac{8\pi G}{c^4}T_{ij}- D_iD_j\varphi
,\label{(5.13)} 
\end{equation} 
where $\varphi\in H^2(M)$ is called
the scalar potential. We infer from (\ref{(5.9)}) that the
conservation laws for (\ref{(5.13)}) are as follows
\begin{equation}
{\rm div}\ (D_iD_j\varphi  + \frac{8\pi
G}{c^4}T_{ij})=0.\label{(5.14)}
\end{equation}
Using the contraction with $g^{ij}$ in (\ref{(5.13)}), we have 
\begin{equation}
R=\frac{8\pi G}{c^4}T +  \Phi,
\end{equation}
where 
$$ T = g^{ij} T_{ij}, \qquad  \Phi= g^{ij}D_iD_j\varphi,$$
represent respectively the energy-momentum density and the scalar potential density. Physically this scalar potential density $\Phi$ represents potential energy caused by the non-uniform distribution of matter in the universe. One important property of this scalar potential is 
\begin{equation}\la{average}
\int_M \Phi \sqrt{-g} dx =0,
\end{equation}
which is due to the integration by parts formula in Theorem~\ref{t2.1}. This formula demonstrates clearly that the negative part of this quantity $\Phi$ represents the dark matter, which produces attraction, and the positive part represents the dark energy, which drives the acceleration of expanding galaxies. We shall address this important issue in the next section.

\subsection{Field equations for closed universe} 
The topological structure  of  closed universe  is  given by 
\begin{equation}
M=S^1\times S^3,\label{4.3}
\end{equation}
where $S^1$ is the time circle and $S^3$ is the 3-dimensional sphere
representing the space. We note that the radius $R$ of $S^3$ depends
on time $t\in S^1$,
$$R=R(t),\ \ \ \ t\in S^1,$$
and the minimum time $t_0$,
$$t_0=\min\limits_tR(t)$$
is the initial time of  the Big Bang.

For a closed universe as (\ref{4.3}), by Theorem~\ref{t4.3}, the
gravitational field equations are in the form
\begin{equation}
\begin{aligned} &R_{ij}-\frac{1}{2}g_{ij}R=-\frac{8\pi
G}{c^4}T_{ij} - D_iD_j\varphi +\alpha D_i\psi_j,\\
&\Delta\psi_j+g^{ik}R_{ij}\psi_k=0,\\
&  D_i\psi_j=D_j\psi_i,
\end{aligned}\label{4.4}
\end{equation}
where $\Delta =D^kD_k, \varphi$ the scalar  potential, $\psi
=(\psi_0,\psi_1,\psi_2,\psi_3)$ the vector potential,   and $\alpha$ is a
constant. The conservation laws of (\ref{4.4}) are as follows
\begin{equation}
\Delta\psi_j= \frac{1}{\alpha}\Delta\left(\frac{\partial\varphi}{\partial
x^j}\right)+\frac{8\pi G}{\alpha c^4}D^kT_{kj}.\label{4.5}
\end{equation}

\section{Interaction in  A Central Gravitational Field}
\subsection{Schwarzschild solution}
We know that the metric of a central gravitational field is in a
diagonal form \cite{atwater}:
\begin{equation}
ds^2=g_{00}c^2dt^2+g_{11}dr^2+r^2(d\theta^2+\sin^2\theta
d\varphi^2),\label{5.1}
\end{equation}
and physically  $g_{00}$ is given by 
\begin{equation}
g_{00}=-\left(1+\frac{2}{c^2}\psi \right),\label{5.2}
\end{equation}
where $\psi$ is the Newton gravitational potential; see among others \cite{atwater}.

If the central matter field has total mass $M$ and radius $r_0$,
then for  $r>r_0$, the metric (\ref{5.1})  is the well known
Schwarzschild solution for the Einstein field equations
(\ref{(5.8)}),  and  is  given by 
\begin{equation}
ds^2= -\left(1-\frac{2MG}{c^2r}\right)c^2dt^2+\frac{dr^2}{(1-\frac{2MG}{c^2r})}
+r^2(d\theta^2+\sin^2\theta d\varphi^2). \label{5.3}
\end{equation}
We derive from (\ref{5.2}) and (\ref{5.3})  the classical
Newton gravitational potential
\begin{equation}
\psi =-\frac{MG}{r}.\label{5.4}
\end{equation}

\subsection{New gravitational interaction model}
We now consider the metric determined by the
new  field equations (\ref{(5.13)}), from which  we  derive a 
gravitational potential formula replacing  (\ref{5.4}).

Equations (\ref{(5.13)}) can be equivalently expressed as
\begin{equation}
R_{ij}=-\frac{8\pi G}{c^4}(T_{ij}-\frac{1}{2}g_{ij}T) -
(D_iD_j\varphi -\frac{1}{2}g_{ij}\Phi ),\label{5.5}
\end{equation}
where
$$T=g^{kl}T_{kl},\ \ \ \ \Phi =g^{kl}D_kD_l\varphi ,$$

For the central matter field with total mass $M$ and radius $r_0$,
by the Schwarzschild  assumption, for $r>r_0$, there exists  no matter, i.e.
\begin{equation}
T_{ij}=0.\label{5.7}
\end{equation}
Therefore the conservation laws of (\ref{5.5}) are
\begin{equation}
\Delta\left(\frac{\partial\varphi}{\partial x^k}\right)=0\qquad \text{  for } 
k=0,1,2,3.\label{5.8}
\end{equation}

The tensors $g_{ij}$ in (\ref{5.1}) are written as
\begin{equation}
\begin{aligned}
&g_{00}=-e^u,&&g_{11}=e^v,&&g_{22}=r^2,&&g_{33}=r^2\sin^2\theta, \\
&  u=u(r),&& v=v(r). 
\end{aligned}\label{5.8-1}
\end{equation}
Noting that the central field is spherically  symmetric, we  assume
that
\begin{eqnarray}
&&\varphi =\varphi (r)\ \text{\rm  is independent of}\ t, \theta ,\varphi
.\label{5.9}\\
&&r\gg\frac{2MG}{c^2}.\label{5.10}
\end{eqnarray}

For the metric (\ref{5.8}), the non-zero components of the Levi-Civita
connection are as follows
\begin{equation}
\label{5.11}
\begin{aligned}
&\Gamma^1_{00}=\frac{1}{2}e^{u-v}u^{\prime}, && 
\Gamma^1_{11}=\frac{1}{2}v^{\prime}, && \Gamma^1_{22}=-re^{-v},\\
&\Gamma^1_{33}=-re^{-v}\sin^2\theta ,&& 
\Gamma^0_{10}=\frac{1}{2}u^{\prime}, &&
\Gamma^2_{12}=\frac{1}{r},\\
&\Gamma^2_{33}=-\sin\theta\cos\theta , &&
\Gamma^3_{13}=\frac{1}{r}, &&
\Gamma^3_{23}=\frac{\cos\theta}{\sin\theta}.
\end{aligned}
\end{equation}
Hence the Ricci tensor 
$$R_{ij}=\frac{\partial\Gamma^k_{ik}}{\partial
x^j}-\frac{\partial\Gamma^k_{ij}}{\partial
x^k}+\Gamma^k_{ir}\Gamma^r_{jk}-\Gamma^k_{ij}\Gamma^r_{kr}
$$ 
are given by 
\begin{equation}
\begin{aligned}
&R_{00}=-e^{u-v}\left[\frac{u^{\prime\prime}}{2}+\frac{u^{\prime}}{r}+\frac{u^{\prime}}{4}(u^{\prime}-v^{\prime})\right],\\
&R_{11}=\frac{u^{\prime\prime}}{2}-\frac{v^{\prime}}{r}+\frac{u^{\prime}}{4}(u^{\prime}-v^{\prime}),\\
&R_{22}=e^{-v}\left[1-e^v+\frac{r}{2}(u^{\prime}-v^{\prime})\right]\\
&R_{33}=R_{22}\sin^2\theta ,\\
& R_{ij}=0\qquad    \forall i\neq j.
\end{aligned}\label{5.12}
\end{equation}
Furthermore, we infer from (\ref{5.8}), (\ref{5.9}) and (\ref{5.11})  that
\begin{equation}
\begin{aligned}
&D_iD_j\varphi -\frac{1}{2}g_{ij}\Phi =0,\ \ \ \ \forall i\neq j, \\
&D_0D_0\varphi -\frac{1}{2}g_{00}\Phi
=\frac{1}{2}e^{u-v}\left[\varphi^{\prime\prime}-\frac{1}{2}(u^{\prime}+v^{\prime}-\frac{4}{r})\varphi^{\prime}\right],\\
&D_1D_1\varphi -\frac{1}{2}g_{11}\Phi
=\frac{1}{2}\left[\varphi^{\prime\prime}-\frac{1}{2}(v^{\prime}+u^{\prime}+\frac{4}{r})\varphi^{\prime}\right],\\
&D_2D_2\varphi -\frac{1}{2}g_{22}\Phi
=-\frac{r^2}{2}e^{-v}\left[\varphi^{\prime\prime}+\frac{1}{2}(u^{\prime}-v^{\prime})\varphi^{\prime}\right],\\
&D_3D_3\varphi -\frac{1}{2}g_{33}\Phi
=\sin^2\theta\left(D_2D_2\varphi
-\frac{1}{2}g_{22}\Phi\right).
\end{aligned}\label{5.13}
\end{equation}
Thus, by (\ref{5.12}) and (\ref{5.13}), the equations
(\ref{5.5}) are as follows
\begin{align}
&u^{\prime\prime}+\frac{2u^{\prime}}{r}+\frac{u^{\prime}}{2}(u^{\prime}-v^{\prime})=   \varphi^{\prime\prime} - \frac{1}{2}(u^{\prime}+v^{\prime}-\frac{4}{r})\varphi^{\prime},\label{5.14}\\
&u^{\prime\prime}-\frac{2v^{\prime}}{r}+\frac{u^{\prime}}{2}(u^{\prime}-v^{\prime})=  - \varphi^{\prime\prime}+ \frac{1}{2}(u^{\prime}+v^{\prime}+\frac{4}{r})\varphi^{\prime},\label{5.15}\\
&u^{\prime}-v^{\prime}+\frac{2}{r}(1-e^v)=
r(\varphi^{\prime\prime}+\frac{1}{2}(u^{\prime}-v^{\prime})\varphi^{\prime}).\label{5.16}
\end{align}

\subsection{Consistency}
We need to consider the existence and uniqueness of solutions of the equations (\ref{5.14})-(\ref{5.16}). 
First, in the vacuum case,  the  classical Einstein equations are in the form
\begin{equation}\la{6.1a}
R_{kk}=0 \qquad \text{ for } k=1, 2, 3,
\end{equation}
two of which are independent. The system contains  two  unknown functions, and therefore for a given initial value (as $u'$ is basic in (\ref{6.1a})):
\begin{equation}\la{6.2a}
u'(r_0)=\sigma_1, \qquad  v(r_0)=\sigma_2, \qquad r_0 >0,
\end{equation}
the problem (\ref{6.1a}) with (\ref{6.2a}) has a unique solution, which is  the Schwarzschild solution 
$$u'=\frac{\varepsilon_2}{r^2}\left(1-\frac{\varepsilon_1}{r}\right)^{-1}, \qquad v=-\ln\left(1-\frac{\varepsilon_1}{r}\right),$$
where 
$$\varepsilon_1 =r_0(1-e^{-\sigma_2}), \qquad \varepsilon_2= r_0^2 \left(1-\frac{\varepsilon_1}{r_0}\right) \sigma_1.$$

Now if we consider the influence of the cosmic microwave background (CMB) for the central fields, then we should add a constant energy density in equations (\ref{6.1a}):
\begin{equation}
\la{6.3a}
(T_{ij})=\left(
\begin{matrix}
-g_{00}\rho & 0 & 0 & 0 \\
0 & 0 & 0 & 0\\
0 & 0 & 0 & 0\\
0 & 0 & 0 & 0\\
\end{matrix}
\right).
\end{equation}
Namely, 
\begin{equation}
R_{00}=\frac{4\pi G}{c^4} g_{00} \rho, \qquad R_{11}=0, \qquad R_{22}=0, \la{6.4a}
\end{equation}
where $\rho$ is the density of the microwave background, whose value is $\rho=4 \times 10^{-31} kg/m^3$. 
Then it is readily to see that the problem (\ref{6.4a}) with (\ref{6.2a})  has no solution.
In fact, the divergence-free equation $D^iT_{ij}=0$  yields that 
$$\Gamma^0_{10} T_{00} =\frac{1}{2} u' \rho=0,$$
which implies that $u'=0$. Hence $R_{00}=0$, a  contradiction to (\ref{6.4a}).
Furthermore, if we regard $\rho$ as an unknown function, then the  equations  (\ref{6.4a}) still have no solutions. 

On the other hand, the new gravitational field  equations (\ref{5.14})-(\ref{5.16}) are solvable for the microwave background as the number of unknowns are the same as the number of independent equations. 

Equations (\ref{5.14})-(\ref{5.16}) have the following equivalent form:
\begin{equation}
\begin{aligned}
& u'' =\left(\frac1r + \frac{u'}2\right) (v'-u') + \frac{\varphi'}r, \\
& \varphi'' = - \frac{1}{r^2}\left( e^v -1\right) +  \frac{1}2 \varphi'v'  + \frac1r u'\\
& v' = - \frac{1}{r} (e^v-1) - \frac{r}{2} \varphi' u',
\end{aligned}\la{6.5a}
\end{equation}
equipped with the following 
initial values:
\begin{equation}
u'(r_0)=\alpha_1,\qquad v(r_0)=\alpha_2, \qquad \varphi'(r_0)=\alpha_3, \qquad r_0 >0.
\la{6.6a}
\end{equation}
It is classical that (\ref{6.5a}) with (\ref{6.6a}) possesses a unique local solution. 
In fact, we can prove that the solution exists for all  $r >r_0$.

\subsection{Gravitational interaction}
We now derive the gravitational interaction formula from the basic model  (\ref{5.14})-(\ref{5.16}) . 

First, we infer from   (\ref{5.14})-(\ref{5.16}) that
\begin{equation}
\begin{aligned}
&u^{\prime}+v^{\prime}=\frac{r\varphi^{\prime\prime}}{1+ \frac{r}{2}\varphi^{\prime}}, \\
&u^{\prime}-v^{\prime}=\frac{1}{1- \frac{r}{2}\varphi^{\prime}}\left[\frac{2}{r}(e^v-1) + r\varphi^{\prime\prime}\right].
\end{aligned}\label{6.8a}
\end{equation}
Consequently, 
\begin{equation}
u^{\prime}=\frac{1}{1-\frac{r}{2}\varphi^{\prime}}\frac{1}{r}(e^v-1)
+ \frac{r\varphi^{\prime\prime}}{1-(\frac{r}{2}\varphi^{\prime})^2}.
\end{equation}
By (\ref{5.2}) and (\ref{5.8}), we have 
\begin{equation}
\psi=\frac{c^2}{2} (e^u-1).\la{6.9a}
\end{equation}
As the interaction force $F$  is given by 
$$F=-m\nabla \psi,$$
it follows from (\ref{6.8a})  and (\ref{6.9a}) that 
\begin{align}
& F=\frac{mc^2}{2} e^u \left[
- \frac{1}{1- \frac{r}{2}\varphi^{\prime}}\frac{1}{r}(e^v-1)
- \frac{r\varphi^{\prime\prime}}{1-(\frac{r}{2}\varphi^{\prime})^2} \right],   \la{6.10a} \\
& \varphi^{\prime\prime}= - e^v R + \frac12 ( u' - v' + \frac4r) \varphi^\prime. \la{6.10b}
\end{align}
Of course, the following energy balance and conservation law hold true as well:
\begin{equation}
R=\frac{8\pi G}{c^4} T + \Phi, \qquad \int_0^\infty e^{u+v} r^2 \Phi dr=0. \la{r-conservation}
\end{equation}
where $R$  is the scalar curvature  and $\Phi = g^{kl} D_k D_l \varphi$. 
Equation  (\ref{6.10b}) is derived by solving  $\varphi''$  using 
(\ref{r-conservation}). Namely, for $r>r_0$ where $T=0$, 
$$
R=\Phi = g^{kl} D_k D_l \varphi =e^{-v}[
-\varphi'' + \frac12(u'-v'+\frac{4}{r})\varphi'].
$$

\subsection{Simplified formulas}
We now consider the region: $r_0< r < r_1$. Physically, we have 
\begin{equation}
|\varphi' |, |\varphi''| <<1.\la{6.11a}
\end{equation}
Hence $u$  and $v$  in (\ref{6.10a}) can be replaced by the Schwarzschild solution:
\begin{equation}
u_0 =\ln\left(1-\frac{\delta}{r}\right), \qquad v_0 =- \ln\left(1-\frac{\delta}{r}\right), \qquad \delta = \frac{2GM}{c^2}.
\la{6.12a}
\end{equation}
As $\delta/r$  is small for $r$ large, by (\ref{6.11a}), the formula (\ref{6.10a})  can be
expressed as 
\begin{equation}
F=\frac{mc^2}2
\left[  -\frac{\delta}{r^2} - \varphi'' r
\right].\la{6.13a}
\end{equation}
This is the interactive force in a central  symmetric field. The first term in the parenthesis is the Newton gravity term, and the added second term $-r\varphi''$ is the scalar potential energy density, representing the dark matter and dark energy.

In addition, replacing $u$  and $v$  in (\ref{6.10b})  by the Schwarzschild solution (\ref{6.12a}), we derive the following approximate formula:
\begin{equation}
\varphi''=\left(\frac2r+ \frac{\delta}{r^2}   \right)\varphi' - R.\la{6.15a}
\end{equation}
Consequently we infer from (\ref{6.13a}) that 
\begin{equation}
F={mMG} \left[  
-\frac{1}{r^2} - \frac1{\delta} \left(2+ \frac{\delta}{r} \right) \varphi' + \frac{Rr}{\delta}\right], \qquad R=\Phi \quad \text{ for } r>r_0. \la{6.16a}
\end{equation}
The first term is the classical Newton gravitation, the second term is the coupling interaction between matter and the scalar potential $\varphi$, and the third term is the interaction generated by the scalar potential energy density $\Phi$.
In this formula, the negative and positive values of each term represent respectively the attracting and repelling forces. 

Integrating (\ref{6.15a}) yields (omitting $e^{-\delta/r}$)
\begin{equation}
\varphi' =-\varepsilon_2 r^2 -r^2 \int r^{-2} R dr, \la{6.1b}
\end{equation}
where $\varepsilon_2$  is a free parameter.  Hence the interaction force $F$  is approximated by 
\begin{equation}
F={mMG} \left[  
-\frac{1}{r^2} +  \left(2+ \frac{\delta}{r} \right) \varepsilon r^2 + \frac{Rr}{\delta} + 
\frac{1}{\delta} \left(2+ \frac{\delta}{r} \right) r^2 \int r^{-2} R dr
\right],\la{f2}
\end{equation}
where $\varepsilon =\varepsilon_2 \delta^{-1}$, $R=\Phi$  for $r>r_0$, and $\delta=2MG/c^2.$  
We note that based on (\ref{r-conservation}), 
for $r> r_0$, $R$  is balanced by  $\Phi$, and the conservation of $\Phi$ suggests that $R$ behaviors like $r^{-2}$  as $r$ sufficiently large. Hence for $r$ large, the second term in the right hand side of (\ref{f2}) must be dominate and positive, indicating the existence of dark energy.

We note that the  scalar curvature is infinite at $r=0$: $R(0)=\infty.$
Also  $R$  contains two free parameters determined by $u'$  and $v$ respectively. Hence if we take a first order approximation as 
\begin{equation}
R=- \varepsilon_1 + \frac{\varepsilon_0}{r} \qquad \text{ for } r_0 < r < r_1=10^{21}km,
\la{6.2b}
\end{equation}
where $\varepsilon_1$  and $\varepsilon_0$  are free yet to be determined parameters. Then we deduce from 
(\ref{6.1b})  and (\ref{6.2b}) that 
$$\varphi' = -\varepsilon \delta r^2 - \varepsilon_1 r + \frac{\varepsilon_0}2.
$$
Therefore,
\begin{equation}
F=mMG\left[-\frac{1}{r^2} -\frac{\varepsilon_0}2 \frac{1}{r} + \varepsilon_1 +(\varepsilon \delta + \varepsilon_1 \delta^{-1})r  
+ 2 \varepsilon  r^2\right].
\end{equation}
Physically  it is natural  to choose 
$$\varepsilon_0 >0, \qquad \varepsilon_1 > 0, \qquad \varepsilon> 0. $$
Also, $\varepsilon_1$   and $2 \varepsilon_2 \delta^{-1} r^2$ are much  smaller than $(\varepsilon \delta +\varepsilon_1 \delta^{-1})r$  for $r \le r_1$. Hence 
\begin{equation}
F=mMG\left[-\frac{1}{r^2} -\frac{k_0}{r} + k_1 r  \right]. \la{f3}
\end{equation}
where $k_0$  and $k_1$  can be estimated using the Rubin law of rotating galaxy and the acceleration of the expanding galaxies:
\begin{equation}
k_0=4\times 10^{-18} km^{-1}, \qquad k_1=10^{-57} km^{-3}.
\end{equation}
We emphasize here that the formula (\ref{f3}) is only a simple  approximation for illustrating  some features of both dark matter and dark energy.

\section{Theory of Dark Matter and Dark Energy}
\subsection{Dark matter and dark energy}
Dark matter and dark energy are two of most remarkable discoveries  in
astronomy  in recent years, and they are introduced to explain the acceleration of the expanding galaxies. In spite of many attempts and theories, the mystery remains.   
As mentioned earlier, this article is an attempt to develop a unified theory for the dark matter and dark energy. 

A strong support to the existence of dark matter is the Rubin law
for galactic  rotational velocity, which amounts to saying that most
stars in spiral galaxies orbit at roughly the same speed. Namely,
the orbital velocity $v(r)$ of the stars located at radius $r$ from
the center of galaxies is almost a constant:
\begin{equation}
v(r)=\text{\rm constant for a given galaxy.}\label{6.1}
\end{equation}
Typical galactic rotation curves \cite{kutner}   are 
illustrated by Figure
\ref{f6.1}(a), where the vertical axis represents the velocity (km/s), and
the horizontal axis is the distance from the galaxy center (extending to
the galaxy radius).

\begin{figure}[hbt]\la{f6.1}
  \centering
  \includegraphics[width=0.9\textwidth]{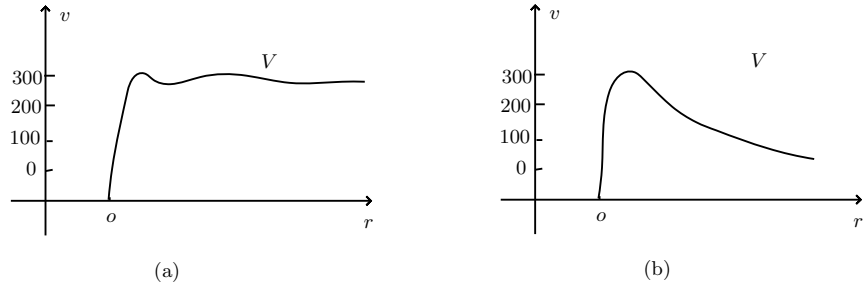}
 \caption{(a) Typical galactic rotation curve by Rubin, and  (b) theoretic curve by the Newton gravitation law.}
\end{figure}

However, observational evidence shows  discrepancies between the
mass of large astronomical objects determined from their
gravitational effects, and the mass calculated from the visible matter
they contain, and Figure~\ref{f6.1} (b) gives a calculated curve. The missing
mass suggests  the presence of dark matter in the universe.

In astronomy and cosmology, dark energy is a hypothetical form of
energy, which  spherically symmetrically permeates all of space
and tends to accelerate the expansion of the galaxies.

The High-Z Supernova Search Team in 1998 and the Supernova
Cosmology Project in 1998 published their observations which reveal
that the expansion of the galaxies is accelerating. In 2011, a
survey of more than $2\times 10^5$ galaxies from Austrian
astronomers confirmed the fact. Thus, the existence of dark energy
is accepted by most astrophysicist.

\subsection{Nature of dark matter and dark energy}
With the new gravitational field equation with the scalar potential energy, and we are now in position to derive the nature of the dark matter and dark energy.
More precisely,  using the revised Newton formula derived from the new field equations:
\begin{equation}
F=mMG\left(-\frac{1}{r^2}-\frac{k_0}{r}+k_1r\right),\label{6.2}
\end{equation}
we  determine an approximation of the  constants $k_0, k_1$, based on the Rubin law and the acceleration of expanding galaxies.

\medskip

First, let $M_r$  be  the total mass in the ball with radius $r$ of the galaxy, and
$V$  be  the constant galactic  rotation  velocity. By the force equilibrium,
we infer  from (\ref{6.2}) that
\begin{equation}
\frac{V^2}{r}=M_rG\left(\frac{1}{r^2}+\frac{k_0}{r}-k_1r\right),\label{6.3}
\end{equation}
which implies that 
\begin{equation}
M_r=\frac{V^2}{G}\frac{r}{1+k_0r-k_1r^3}.\label{6.4}
\end{equation}
This matches  the  observed mass distribution formula of the galaxy, which can explain
the Rubin law (\ref{6.3}).

\medskip

Second, if we use the classical  Newton formula
$$F=-\frac{mMG}{r^2},$$
to calculate the galactic  rotational velocity $v_r$, then we have
\begin{equation}
\frac{v^2_r}{r}=\frac{M_rG}{r^2}.\label{6.5}
\end{equation}
Inserting (\ref{6.4}) into (\ref{6.5}) implies 
\begin{equation}
v_r=\frac{V}{\sqrt{1+k_0r-k_1r^3}}.\label{6.6}
\end{equation}
As $1\gg k_0\gg k_1$, (\ref{6.6}) can approximatively written as
$$v_r=V \left(1-\frac{1}{2}k_0r + \frac14 k_0^2 r^2 \right),$$
which is consistent with the theoretic rotational curve as illustrated
by Figure \ref{f6.1}(b). It implies that the distribution formula
(\ref{6.4}) can be used as a test for the revised gravitational
field equations.

\medskip

Third, we now determine the constants $k_0$ and $k_1$ in (\ref{6.2}).
According to astronomic data, the average mass $M_{r_1}$ and radius
$r_1$ of galaxies is about
\begin{equation}
\begin{aligned}
&M_{r_1}=10^{11}M_{\odot}\simeq 2 \times 10^{41}{\rm kg},\\
&r_1=10^4\sim 10^5pc\simeq 10^{18}{\rm km},
\end{aligned}\label{6.7}
\end{equation}
where $M_{\odot}$  is the mass of the Sun.

Taking $V=300{\rm km/s}$, then we have
\begin{equation}
\frac{V^2}{G}=8 \times 10^{23}{\rm kg/km}.\label{6.8}
\end{equation}
Based on physical considerations,
\begin{equation}
k_0\gg k_1r_1\ \ \ \ (r_1\ \text{\rm as in (\ref{6.7})})\label{6.9}
\end{equation}
By (\ref{6.7})-(\ref{6.9}), we deduce from (\ref{6.4}) that
\begin{equation}
k_0=\frac{V^2}{G}\frac{1}{M_{r_1}}-\frac{1}{r_1}=4 \times 10^{-18}{\rm
km}^{-2}.\label{6.10}
\end{equation}

Now we consider the constant $k_1$. Due to the accelerating expansion of
galaxies, the interaction force between two clusters of galaxies
is repelling, i.e. for (\ref{6.2}),
$$F\geq 0,\ \ \ \ r\geq \bar r,$$
where $\bar r$ is the average distance between two galactic clusters. It is
estimated that
$$\bar r=10^8{\rm pc}\simeq 10^{20}\sim 10^{21}{\rm km}.$$
We take
\begin{equation}
\bar r=\frac{1}{\sqrt2} \times 10^{20}{\rm km}\label{6.11}
\end{equation}
as the distance at which $F=0$. Namely, 
$$k_1 \bar r -\frac{k_0}{\bar r}-\frac{1}{\bar r^2}=0.$$
Hence we derive from (\ref{6.10}) and (\ref{6.11}) that 
$$k_1=k_0\bar r^{-2}=10^{-57}{\rm km}^{-3}.$$
Thus, the constants $k_0$ and $k_1$ are estimated by
\begin{equation}
 k_0=4  \times 10^{-18}{\rm km}^{-2},\qquad 
k_1=10^{-57}{\rm km}^{-3}.
\label{6.12}
\end{equation}

In summary, for the formula (\ref{6.2}) with (\ref{6.12}), if
the matter distribution $M_r$ is in the form
\begin{equation}
M_r=\frac{V^2}{G}\frac{r}{1+k_0r},\label{6.13}
\end{equation}
then the Rubin law holds true. In particular, the mass $\tilde{M}$
generated by  the  revised gravitation is
$$\tilde{M}=M_T-M_{r_1}=\frac{V^2}{G}r_1 -\frac{V^2}{G}\frac{r_1}{1+k_0r_1},\
\ \ \ r_1\ \text{\rm as in (\ref{6.7})},
$$ 
where $M_T=V^2r_1/G$ is the
theoretic value of total mass. Hence
$$\frac{\tilde{M}}{M_T}=\frac{k_0r_1}{1+k_0r_1}=\frac{4}{5}.$$
Namely, the revised gravitational mass $\tilde{M}$ is four times of 
the visible matter $M_{r_1}=M_T-\tilde{M}$. Thus, it gives an alternative
explanation for the dark matter.

In addition, the formula (\ref{6.2}) with (\ref{6.12}) also
shows that for a central field with mass $M$,   an object at $r>\bar r$
($\bar r$ as in (\ref{6.11})) will be exerted a repelling force,
resulting the acceleration of expanding galaxies at $r> \bar r$. 

Thus  the new gravitational formula (\ref{6.2}) provides 
a unified  explanation of dark matter and dark energy.

\subsection{Effects of non-homogeneity}
In this section, we prove that  if the matter is homogeneously 
distributed in the universe, then  the scalar potential $\varphi$ is a constant, and consequently the 
scalar potential energy density is identically zero:  $\Phi\equiv 0$. 

It is known that the metric for an isotropic and homogeneous universe is given by 
\begin{equation}
ds^2=- c^2dt^2+a^2\left[ \frac{dr^2}{1-kr^2} +r^2(d\theta^2+\sin^2\theta
d\varphi^2)\right],\la{7.1}
\end{equation}
which is called the Friedmann-Lema\^itre-Robertson-Walker (FLRW) metric, where    the 
scale  factor $a=a(t)$ represents   the cosmological radius, and $k$ takes one of the three numbers: $-1, 0, 1$.

For the FLRW metric (\ref{7.1}), the energy-momentum tensor $\{T_{ij}\}$ is given by 
\begin{equation}
T_{ij} = \text{diag} (\rho c^2, g_{11} p, g_{22} p, g_{33}p), \la{7.2}
\end{equation}
where $\rho$  is the mass density, $p$ is pressure, and 
$$g_{11}=\frac{a^2}{1-k r^2}, \qquad g_{22}= a^2 r^2, g_{33}= a^2 r^2 \sin^2 \theta.$$
By (\ref{7.1})  and (\ref{7.2}), the Einstein field equations (\ref{1.1}) are reduced to two equations:
\begin{align}
& a''= -\frac{4 \pi G}{3} \left(\rho + \frac{3 p}{c^2}\right) a,  \la{7.3}  \\
& a a'' + 2 (a')^2 + 2 k c^2 = 4 \pi G \left(  \rho- \frac{p}{c^2}    \right)a^2, \la{7.4}
\end{align}
and the conservation law $\text{ div } T_{ij}=0$ gives 
\begin{equation}
\frac{d\rho}{dt} + \frac{3}{R} \frac{dR}{dt} \left( \rho + \frac{p}{c^2}\right)=0.  \la{7.5}
\end{equation}
Then only two of the above three equations (\ref{7.3})-(\ref{7.5})  are independent, and are called the 
Friedman equations.

On the other hand, for the metric  (\ref{7.1})  with (\ref{7.2}), 
the  new gravitational field equations (\ref{1.3}) with scalar potential are reduced 
to 
\begin{align}
& a''= -\frac{4 \pi G}{3} \left(\rho + \frac{3 p}{c^2}\right) a  + \frac{1}{6} \varphi'' a - \frac12 a' \varphi',  \la{7.6}  \\
& \frac{a''}{a} + 2 \left(\frac{a'}{a} \right)^2 + \frac{2 k c^2}{a^2} = 4 \pi G \left(  \rho- \frac{p}{c^2}    \right)  
-\frac{\varphi''}2 - \frac{\varphi'}{2} \frac{a'}{a}, \la{7.7}
\end{align}
and the conservation equation 
\begin{equation}
\varphi''' + 3  \frac{a'}{a} \varphi'' = 8 \pi G \left( \rho' + 3  \frac{a'}{a} \rho + 3  \frac{a'}{a} \frac{p}{c^2}\right). 
\la{7.8}
\end{equation}
Again two of the above three equations (\ref{7.6})-(\ref{7.8})  are independent. It follows 
from (\ref{7.6})  and (\ref{7.7})  that 
\begin{equation}
(a')^2 = \frac{8\pi G}{3} a^2 \rho -\frac13 \varphi'' a^2 - k c^2.\la{7.9}
\end{equation}
In the following we shall prove that 
\begin{equation}
\nabla \varphi = 0,  \la{7.10}
\end{equation}
Namely, $\varphi=$constant.
In fact, let $\varphi''$  and $\rho$  have the form:
\begin{equation}
\rho=\frac{\theta}{a^3}, \qquad \varphi'' = \frac{\psi}{a^3}.\la{7.11}
\end{equation}
Inserting (\ref{7.11}) in (\ref{7.6}), (\ref{7.8})  and (\ref{7.9}), we arrive at 
\begin{align}
& a''= -\frac{4 \pi G}{3}\frac{\theta}{a^2}     + \frac{1}{6} \frac{\psi}{a^2} 
 - \frac{4\pi G}{c^2} \theta a   - \frac{1}{2} a' \varphi',  \la{7.12}  \\
&(a')^2 -\frac{8 \pi G}{3}\frac{\theta}{a} 
 + \frac13 \frac{\psi}{a} =
-k c^2 , \la{7.13}\\
& \psi'=8\pi G \theta' + 24 \pi G a^2 a' p/c^2. \la{7.14}
\end{align}
Multiplying both sides of (\ref{7.12})  by $a'$ we obtain 
\begin{equation}
\frac12 \frac{d}{dt} 
\left[ (a')^2 - \frac{8\pi G}{3}\frac{\theta}{a} + \frac13 \frac{\psi}{a}\right] 
+ \frac{4\pi G}{3} \frac{\theta'}{a} - \frac16 \frac{\psi'}{a}   =
- \frac{4\pi G}{c^2} p a a' - \frac12 (a')^2 \varphi'. \la{7.15}
\end{equation}
It follows then from (\ref{7.13})-(\ref{7.15}) that 
$$\frac12 (a')^2 \varphi'=0,$$
which implies that (\ref{7.10}) holds true. 

The conclusion (\ref{7.10}) indicates that if the universe is in the homogeneous state, 
then the scalar potential energy density $\frac{c^4}{8\pi G}\Phi$ is identically zero: $\Phi\equiv 0$. This fact again 
demonstrates that $\varphi$ characterizes  the non-uniform distribution of matter in the universe. 

\section{Conclusions}
We have discovered new gravitational field equations (\ref{1.3}) with scalar potential under the postulate that the energy momentum tensor $T_{ij}$ needs not to be divergence-free due to the presence of dark energy and dark matter:
$$
R_{ij}-\frac{1}{2}g_{ij}R=-\frac{8\pi G}{c^4}T_{ij}- D_iD_j\varphi, 
$$

With the new field equations, we have obtained the following physical conclusions:

\medskip

{\sc First,} gravitation is now described by the Riemannian metric
$g_{ij}$,  the scalar potential $\varphi$  and their interactions, unified by the new gravitational field equations (\ref{1.3}).

\medskip

{\sc Second}, associated with the scalar potential $\varphi$  is the scalar potential energy density $\frac{c^4}{8\pi G}\Phi$, which  represents a new type of  energy/force  caused by the non-uniform distribution of matter in the universe.  This scalar potential energy density  varies as  the galaxies move and matter of the universe redistributes. Like gravity, it affects every part of the universe as a field. 

This scalar potential energy density $\frac{c^4}{8\pi G}\Phi$  consists of both positive and negative energies.  
The negative part of this potential energy density  produces attraction, and the positive part produces repelling force. Also, this scalar energy density is conserved with mean zero:
$$\int_M \Phi dM =0.$$

\medskip

{\sc Third},  using the new field equations, for a spherically symmetric central field with mass $M$  and radius $r_0$, the force  exerted on an object of mass $m$ at distance $r$ is given by (see  (\ref{6.16a})):
$$
F={mMG} \left[  
-\frac{1}{r^2} - \frac1{\delta} \left(2+ \frac{\delta}{r} \right) \varphi' + \frac{Rr}{\delta}\right], \qquad R=\Phi \quad \text{ for } r>r_0.
$$
where $\delta = 2MG/c^2$. 

\medskip

{\sc Fourth,}  the sum $\varepsilon=\varepsilon_1 + \varepsilon_2$ of this new potential energy density 
$$\varepsilon_1=\frac{c^4}{8\pi G} \Phi$$ and the coupling energy   between the energy-momentum tensor $T_{ij}$ and the scalar potential field $\varphi$
$$\varepsilon_2=-\frac{c^4}{8\pi G} \left( \frac{2}{r} + \frac{2MG}{c^2r^2}  \right) \frac{d\varphi}{dr},$$  
gives rise to a new unified theory for dark matter and dark energy: 
The negative part of $\varepsilon$  represents the dark matter, which produces attraction, and the positive part represents the dark energy, which drives the acceleration of expanding galaxies.

\medskip

{\sc  Fifth}, the scalar curvature $R$ of space-time obeys:
$$ R= \frac{ 8\pi G}{c^3} T + \Phi.$$
Consequently, when there is no normal matter present (with $T=0$), the curvature $R$ of space-time is balanced by $R=\Phi$. Therefore, there is no real vacuum in the universe.

\bibliographystyle{siam}

\end{document}